\newif\iffullversion
\fullversiontrue

\newif\ifdraftversion

\newif\ifanonversion

\newif\ifacmversion

\newcommand{\thepdftitle}{Enhancing Privacy, Neglecting Harms: An Analysis of Real-World Digital Privacy Incidents}

\PassOptionsToPackage{dvipsnames,table}{xcolor}

\ifacmversion
\ifanonversion
\documentclass[sigconf,anonymous,review]{acmart}
\else
\documentclass[sigconf]{acmart}
\fi
\usepackage{popets}
\else
\documentclass[letterpaper,twocolumn,10pt]{article}
\usepackage{usenix}
\fi

\ifacmversion
\setcopyright{popets}
\copyrightyear{YYYY}

\acmYear{YYYY}
\acmVolume{YYYY}
\acmNumber{X}
\acmDOI{XXXXXXX.XXXXXXX}
\acmISBN{}
\acmConference{Proceedings on Privacy Enhancing Technologies}
\fi

\usepackage[utf8]{inputenc}

\usepackage{amsmath}
\usepackage{amsthm}
\usepackage{amsfonts}
\usepackage{amssymb}
\usepackage{nicefrac}
\usepackage{subcaption}

\usepackage{longtable}
\usepackage{booktabs}
\usepackage{multirow}

\usepackage{xurl}

\usepackage[normalem]{ulem}
\usepackage{xifthen}
\usepackage{xspace}
\usepackage[inline]{enumitem} 

\newlist{enuminline}{enumerate*}{1}
\setlist[enuminline,1]{label=(\arabic*)}
\setlist[enumerate,1]{itemsep=0.5em}

\usepackage{dashbox}

\usepackage{tikz}
\usetikzlibrary{calc,backgrounds,shapes.geometric,positioning,decorations,decorations.pathreplacing,arrows.meta,fit}
\usepackage{pgfplotstable}
\usepackage{makecell}

\PassOptionsToPackage{colorlinks=true,linkcolor=black,urlcolor=black,citecolor=black,pdftitle=\thepdftitle}{hyperref}
\usepackage{hyperref}
\usepackage[capitalise,nameinlink]{cleveref}

\ifdraftversion
\newcommand{\sv}[1]{{\color{blue}{\textsf{\textbf{sv:} #1}}}}

\newcommand{\TODO}[1]{%
	\ifthenelse{\isempty{#1}}%
		{{\color{red}\textsf{\textbf{TODO}}}}%
		{{\color{red}\textsf{\textbf{TODO:} #1}}}%
}

\newcommand{\old}[1]{{\color{RawSienna}\ifmmode\text{\sout{\ensuremath{#1}}}\else\sout{#1}\fi}}

\else
\fi

\makeatletter
\newcommand{\fullonly}{\relax\iffullversion\expandafter\@firstofone\else\expandafter\@gobble\fi}
\newcommand{\fullelse}{\relax\iffullversion\expandafter\@firstoftwo\else\expandafter\@secondoftwo\fi}
\newcommand{\nonfullonly}{\relax\iffullversion\expandafter\@gobble\else\expandafter\@firstofone\fi}
\makeatother

\makeatletter
\newcommand{\acmonly}{\relax\ifacmversion\expandafter\@firstofone\else\expandafter\@gobble\fi}
\newcommand{\acmelse}{\relax\ifacmversion\expandafter\@firstoftwo\else\expandafter\@secondoftwo\fi}
\newcommand{\nonacmonly}{\relax\ifacmversion\expandafter\@gobble\else\expandafter\@secondofone\fi}
\makeatother

\newcommand{\parabf}[1]{\vspace{2mm}\noindent\textbf{#1}} 
\newcommand{\parait}[1]{\vspace{1mm}\noindent\textit{#1}} 

\iffullversion

\else

\fi

\DeclareMathAlphabet{\mathsc}{OT1}{cmr}{m}{sc}

\makeatletter
\newcommand\gobblepars{%
    \@ifnextchar\par%
        {\expandafter\gobblepars\@gobble}%
        {}}
\makeatother

\theoremstyle{plain}
\newtheorem{theorem}{Theorem}[section]

\theoremstyle{definition}
\newtheorem{definition}[theorem]{Definition}
\ifacmversion
\else
\newtheorem{example}[theorem]{Example}
\fi

\crefname{figure}{Fig.}{Figs.}
\Crefname{figure}{Fig.}{Figs.}
\crefformat{section}{#2\S#1#3}
\Crefformat{section}{#2\S#1#3}
\crefformat{subsection}{#2\S#1#3}
\Crefformat{subsection}{#2\S#1#3}
\crefformat{subsubsection}{#2\S#1#3}
\Crefformat{subsubsection}{#2\S#1#3}
\crefname{appendix}{App.}{Apps.}
\Crefname{appendix}{App.}{Apps.}

\makeatletter
\AddToHook{cmd/appendix/before}{\def\cref@section@alias{appendix}\def\cref@subsection@alias{appendix}}
\makeatother

\newcommand{\numarticles}{178\xspace}
\newcommand{\numincidents}{257\xspace}
\newcommand{\numattacks}{341\xspace}
\newcommand{\numsystems}{94\xspace}

\definecolor{darkblue}{rgb}{0,0,0.5}
\definecolor{darkgreen}{rgb}{0,0.5,0}
\definecolor{darkred}{rgb}{0.5,0,0}

\definecolor{tudred}{RGB}{230,0,26}
\definecolor{tudgreen}{RGB}{153,192,0}
\definecolor{tudorange}{RGB}{245,163,0}
\definecolor{tudblue}{RGB}{0,104,157}
\definecolor{tudpurple}{RGB}{149,17,105}
\definecolor{tudbrown}{RGB}{169,73,19}

\tikzstyle{entity} = [circle,draw=#1,fill=#1]
\tikzstyle{indiv} = [entity=black]
\tikzstyle{person} = [entity=cyan]
\tikzstyle{orga} = [rectangle, draw=orange, fill=orange, minimum width=3mm, minimum height=3mm]
\tikzstyle{gov} = [isosceles triangle, draw=darkgreen, fill=darkgreen, rotate=90, minimum width=3mm,isosceles triangle apex angle=60, scale=0.8]

\begin{document}
\ifacmversion
\else
\title{\Large \bf Enhancing Privacy, Neglecting Harms: \\
An Analysis of Real-World Digital Privacy Incidents} 
\date{}
\fi

\ifanonversion
\author{
{\rm Anonymous Author(s)}\\
{Anonymous Institution(s)}
}
\else
\ifacmversion
\author{Shannon Veitch}
\affiliation{\institution{ETH Zurich}}
\email{sveitch@ethz.ch}

\author{C. Shem}
\affiliation{\institution{University of Alberta}}
\email{c.shem@ualberta.ca}

\author{Lena Csomor}
\affiliation{\institution{ETH Zurich}}
\email{lcsomor@student.ethz.ch}

\author{Oleksandr Dudiy}
\affiliation{\institution{University of Alberta}}
\email{dudiy@ualberta.ca}


\author{Naone Kim}
\affiliation{\institution{University of Alberta}}
\email{naone@ualberta.ca}

\author{Khoi Le}
\affiliation{\institution{University of Alberta}}
\email{phuocngu@ualberta.ca}

\author{Lina Saha}
\affiliation{\institution{University of Alberta}}
\email{linaaman@ualberta.ca}

\author{Alexander Viand}
\affiliation{\institution{Belmont Labs}}
\email{alexander.viand@gmail.com}

\author{Anwar Hithnawi}
\affiliation{\institution{University of Toronto}}
\email{ahithnawi@cs.toronto.edu}

\author{Bailey Kacsmar}
\affiliation{\institution{University of Alberta}}
\email{kacsmar@ualberta.ca}

\ifacmversion
\renewcommand{\shortauthors}{Veitch et al.}
\fi
\else
\author{%
{\rm Shannon Veitch$^1$} \quad
{\rm C. Shem$^2$}  \quad
{\rm Lena Csomor} \quad
{\rm Oleksandr Dudiy$^3$} \quad
{\rm Naone Kim$^3$} \\[0.5ex]
{\rm Khoi Le$^3$} \quad
{\rm Lina Saha$^3$} \quad
{\rm Alexander Viand$^{4,5}$} \quad
{\rm Anwar Hithnawi$^6$} \quad
{\rm Bailey Kacsmar$^3$}
\\[1ex]
$^1$ETH Zurich \quad
$^2$University of Waterloo \quad
$^3$University of Alberta \\[0.3ex]
$^4$Belfort Labs \quad
$^5$University of Cambridge \quad
$^6$University of Toronto
}
\fi
\fi

\ifacmversion
\else
\maketitle
\fi

\begin{abstract}
\ifdraftversion
\fi


Privacy-enhancing technologies 
(PETs) have emerged as a technical means for providing individuals with greater control over their information. Yet despite the growing deployment of PETs, people continue to experience privacy harms. 
In this work, we revisit our understanding of privacy incidents and the realities of those experiencing privacy harms, to assess whether the goals and abilities of PETs are misaligned with the harms people face.

For our study, we collect news articles 
that correspond to a sample of \numincidents real-world privacy incidents. 
We employ content analysis over the articles to 
develop a new information flow 
model that encompasses the complexity of data flows and their relation to resulting harms.
We demonstrate that our model captures both established and novel aspects of privacy incidents and their mitigations. 
In particular, it captures why consent is often insufficient to prevent privacy violations, how harms emerge from complex interactions among multiple entities and actions, and reveals a flaw in our understanding of PETs: 
a focus on enabling functionalities still permits the harms inherent in those functionalities.
Moreover, we find that the entities best positioned to implement harm-preventing measures for the incidents in our sample are the least incentivized to do so.
Overall, our model and analysis identify limitations of privacy technology research for harm prevention and further identifies paths for transforming how we approach the advancement of these technologies. 


\noindent
\textbf{Content Warning:} This paper includes the discussion or mention of applications used to create non-consensual explicit imagery, incidents of intimate partner violence, and suicide.
\end{abstract}

\section{Introduction}
New technologies correspond to new ways for data to be collected and used. 
Each of smart devices, social media, and large language models have had their own history of creating data flows that lead to privacy violations~\cite{booth2025smart,kelley2023there,obar2018biggest}.
In response to these new privacy violations, privacy proponents strive to develop systems and tools for users to mitigate the loss of control over their information.

Privacy-enhancing technologies (PETs) are a class of technical measures 
promoted as solutions for protecting individuals' privacy 
and ensuring compliance in data 
use~\cite{usostp,ico}.
In recent years we have seen an uptake in deployments of PETs by major tech companies 
\ifacmversion
\else 
\cite{apple-fhe,meta-pm,google-pjc}.
\fi
However, harms resulting from privacy incidents continue to be experienced by people.
In 2021, a media organization purchased app data from a data broker, which was correlated with a Wisconsin priest's mobile phone.
They determined that he had visited gay bars while using Grindr, which ultimately led to his resignation~\cite{grindr-priest}.
At the University of Melbourne, Wi-Fi location data was used as evidence in misconduct trials against students attending pro-Palestine protests~\cite{uni-protest}.
Meanwhile, an online gamer used the IP address of another player to find his physical address and initiate a barrage of harassment~\cite{cod-doxxing}.

Somewhere between theory and practice, PETs are faltering when it comes to harm prevention.
Towards a bottom-up approach, the concept of \emph{privacy by design} suggests 
incorporating PETs from early design stages~\cite{cavoukian,Hoepman14,GTD15,privbydesign}.
Although these principles provide a basis for system designers to work with, there is no clear connection between any particular strategy and the harms that they might prevent.
Indeed, much of the development of privacy technologies is not motivated by harm prevention, but by regulatory pressures and desires to enable new functionality while maintaining user trust~\cite{de2025pets}.

In light of these observations,
we build on the position that PETs should be motivated by harm prevention~\cite{troncoso25,ristenpart26}.
Past work has focused on developing a systematic understanding of the (privacy) harms introduced by digital technologies~\cite{CHI:LYVTFD24,SP:TABBBC21,AIES:SRHMRNY23}, arguing that privacy law should focus on the harms arising from technologies and data use~\cite{NWUL:solove23,CLR:AngCal23}, as well as defining and classifying privacy harms \cite{calo,harms}.
But while all of this work offers language to express and reason about privacy incidents and harms, it is largely disconnected from the processes of PETs development and privacy engineering (with some exceptions, e.g., \cite{FAT:KOTG20,EC:ACDJ26}).
Thus, we start by taking a step back to test assumptions about the causes of privacy violations through
analyzing real-world privacy incidents.
We reformulate the conceptual representation of privacy incidents to focus on the concrete harms they bring to individuals.
Under this harm-informed lens, we assess each incident for the potential role of PETs and legal instruments in mitigating the resulting harms.
Further, we report on the role of stakeholder incentives and interests relevant to the deployment of PETs in the context of the incident.
The above attributes of our work combine to address the following research questions:

\begin{itemize}[leftmargin=30pt]
\item[\textbf{RQ1:}] How are privacy incidents that involve digital technologies causing harm?
\item[\textbf{RQ2:}] By what means (regulatory or technological) can we mitigate these harms?
\item[\textbf{RQ3:}] What incentive structures for decision makers must be addressed in order to effectively deploy harm-preventing solutions?
\end{itemize}

RQ1 exposes the range of harmful scenarios that are not currently addressed by PETs, while RQ2 and RQ3 uncover \emph{why} PETs fail to prevent harms.
To answer these questions, we use news media as a sample of real-world privacy incidents.
We develop and apply our model of privacy-based attacks to systematically capture the ways in which technological systems lead to privacy harms.
Our qualitative analysis of \numincidents privacy incidents elucidates the relationships between threat actors and privacy harms 
as well as the complexity of information flows in the real world.
We further examine the incidents with respect to potential harm mitigation while preserving the functionality of the system.

\parabf{Contributions.}
We introduce a new approach to modeling privacy incidents that explicitly connects violations to the resulting harms, and enables a more holistic analysis of harm-preventative measures.
%
Our findings provide a comprehensive view of how privacy-based attacks operate in three regions and the fundamental obstacles to mitigation.
We show that many attacks targeting individuals require few resources and are technologically unsophisticated, while those that apply more power (e.g., relying on broad surveillance) 
act in an untargeted manner or towards subgroups of people (e.g., to discriminate).
Additionally, our new modeling of privacy incidents
still captures established issues with certain approaches to preventing privacy violations, like consent, while also bringing forth new complexities of information flows not previously captured.
Our harm-centric approach identifies a fundamental obstacle: the functionality enabled by certain technological systems itself implies harms. Thus a view of PETs as functionality- and privacy-enhancing is misaligned with the goal of harm prevention.
Even when this conflict does not apply, 
challenges persist due to a range of disincentives for deploying interventions (beyond economic and usability concerns), upon those who have the capacity to enact change.
This observation ultimately surfaces a need for regulatory changes to shift priorities of users and service providers to prevent harms. 

\section{Related Work}
\parabf{Privacy Harms and Violations.}
Prior work has modeled privacy incidents and
developed frameworks for the identification of issues during threat assessment~\cite{linddun,mitre,CHI:DRG23}.
However, these works focus primarily on privacy violations without the resulting harms, or are limited to a single threat actor.
With our methodology and consequent model, we overcome these limitations, as we discuss below. 

Consider, for example, how other information flow models have proven useful, namely the contextual integrity (CI) framework~\cite{CI}.
CI focuses on information flows between two parties, 
and describes the information type and the conditions under which an information flow occurs.
By relying on privacy norms, CI identifies which flows are appropriate, and examines how they affect societal interests.
CI's use of \emph{inappropriate} information flow is akin to our use of \emph{privacy violation}.
While the CI framing 
enables precise statements about what constitutes a privacy incident, it provides little information about the ramifications of the incident.
Our model captures information flows that lead to the privacy violation 
\emph{and} the downstream harm(s).

Concurrently, MITRE's privacy threat modeling framework~\cite{mitre} 
describes privacy incidents as a set of privacy actions and relies on Solove's taxonomy of privacy~\cite{solove2005} to classify harms.
These harms are generally more abstract than the concrete harms that we consider, and 
closer in nature to our definition of privacy violations or CI's inappropriate flows.
Our model captures temporal information to make clear how each action relates to one another (and ultimately, the harm).



Finally, the MAP privacy threat modeling framework~\cite{CHI:DRG23} captures the harms resulting from privacy threats, but it does so only with respect to a single threat actor. 
While the model recognizes that threat actors may have different intentions, the framework is not able to capture relationships between multiple entities whose actions combine to cause harms (e.g., one entity using another's technology to cause harm).
This is particularly important when evaluating potential interventions as additional entities have different means of deploying harm-preventing measures. 


Beyond frameworks for privacy threat modeling,
works that apply a privacy theory (e.g., CI~\cite{nissenbaum04,CI} or harm classification~\cite{solove2005,calo}) to hypothetical or measurable incidents typically either measure how well the theory captures the incidences or focuses on what is left out by the theory~\cite{woodruff2014would,wijesekera2015android,ASMRF18,BS23}.
However, not all works in the theory domain focus on modeling or abstraction; others narrow their scope by compiling technology-specific privacy incidents in specific domains or technologies, such as artificial intelligence or robotics~\cite{CHI:LYVTFD24,elderlyrobots,CHI:WWSM23,SP:TABBBC21,CHI:GKH26}. While these offer specialized understanding of specific domains 
our modeling of privacy-based attacks emphasizes the breadth of technologies and services that result in privacy harms.

\parabf{Mitigations and Barriers to Adoption.}
Furthermore, through our focus on the connection between violations and the consequences to the individuals we can identify where and what mitigation efforts are needed to mitigate or prevent these incidents.
That is, our derived system is flexible in describing attacks, connected to harms, and supports an analysis of potential mitigations, which we demonstrate through our secondary analysis. 
We broaden the scope of prior work and leverage expertise in privacy technologies to evaluate a wider range of harm mitigations~\cite{AIES:SRHMRNY23,CHI:SBDDG23,CHI:WWSM23,CHI:FPMLRD16,SP:TABBBC21,CHI:WK24,FAT:KOTG20}, exploring the failures and potential of PETs for harm prevention.

Studies on barriers to adoption of PETs have broadly focused on the usability of technical systems~\cite{FH04,USENIX:WhiTyg99} and have identified barriers including social factors~\cite{networkeffect,CHI:GFF06}, trust in the system~\cite{SOUPS:ABHLG15}, and lack of awareness~\cite{PETS:RenVolRen14}. Previous work also analyzed the challenges faced by users towards adopting specific PETs~\cite{PoPETS:NWCK20,PoPETS:SSYACS21,PoPETS:HarPapRan20}.
Ultimately, it is well established that deploying privacy protections entails costs.
We broaden this line of work to interventions aimed at preventing harms by assessing how technical and regulatory interventions impact both users and system providers.

\parabf{Analyses of News Media.}
Finally, we note past empirical analyses of privacy incidents drawn from news media and case studies.
Sentiment analyses of media has found the overall sentiment in privacy media to be negative~\cite{PST:SheAjmSta17,PoPETS:MirVilPop24} and in terms of topic coverage, there is variance across different regions~\cite{PoPETS:MirVilPop24,SanYas21}. 
While such works have given insights into privacy reporting ecosystems, our work focuses on the incidents being reported, rather than the reporting itself.

\section{Methodology Overview}
We first employ an unobtrusive research method, namely content analysis where news articles are our texts, to develop a new understanding and model of technology-based privacy harms and how they materialize~\cite{glaser1998grounded,kellehear2020unobtrusive}.
Our new model of privacy-based attacks is iteratively derived from our analysis of the articles. Our model provides a 
systematic way of describing and analyzing privacy incidents.
As a secondary analysis, we assess how the harms identified during the content analysis of the news articles can be mitigated, through privacy technologies or regulatory means.

\ifacmversion
\else 
\parabf{Scope.}
\fi
We begin by defining the scope of incidents relevant to our study. To account for different understandings of privacy across geographic regions and contexts we derive our definition of a privacy violation.
We merge notions from Westin's~\cite{westin67} definition of privacy as control over personal information along with ideas from contextual integrity~\cite{nissenbaum04}
to produce our treatment of privacy incidents as information flows, where at some point in the flow an individual loses control over their data.
We merge these existing theories to define \emph{privacy violations}. 
\begin{definition}[Privacy Violation]\label{def:privviol}
    An action or inaction that causes an entity to lose control about how, when, or to what extent personal information about it is communicated to others, outside the scope of accepted practices.
\end{definition}

Since our focus is on privacy violations which result in a concrete harm, we adopt the following definition of harm, whereby impairment or disruption, we include any negative impact on individuals, groups of individuals, or society, as can be captured by existing taxonomies of harms~\cite{harms}.
In our analysis, we only consider events where it is explicitly stated that a harm occurred and where it is clear that the harm is a consequence of the privacy violation.

\begin{definition}[Harm \cite{cyberpeace}]\label{def:harm}
   An impairment or disruption of an entity's capacity or ability to function and exist as it otherwise would have in its usual context.
\end{definition}

Finally, we restrict our evaluation to incidents related to digital technology, and
to events where digital systems function as intended.
This excludes security incidents, where systems may reveal information due to software bugs, malware, or data breaches.
While we recognize that privacy technologies can reduce the harms resulting from such incidents,
limiting inclusion to \emph{non-breach} events assures we do not overwhelm our model~\cite{nonbreach,mitre}.
Collectively, these constraints ensure a scope specific enough to acquire data which addresses our research questions without over-collection of adjacent samples.

\ifacmversion
\else
\parabf{Data Collection.}
\fi
To capture a range of incidents within our scope, we collect (online) newspaper articles from a selection of sources. 
News coverage 
accounts for a variety of topics, and includes incidents with and without legal consequences,
providing insight into the ways in which privacy harms emerge.

We selected sources from three regions: United States (US), United Kingdom (UK), and Australia (AU).
We selected two news sources per region based on their popularity in the region and their likelihood to include coverage of technology-related news.
\ifacmversion\else We \fi
searched for articles published within a particular time period that contained relevant keywords.
Then we applied a set of exclusion and inclusion criteria to ensure that each article contained a privacy violation resulting in a concrete harm, according to Def.~\ref{def:privviol} and \ref{def:harm}.
The details and results of the data collection process are reported in \Cref{sec:collection}.

\ifacmversion
\else
\parabf{Article Analysis.}
\fi
We analyzed all incidents found in the articles using an inductive approach to allow a common structure to emerge. 
Through the process, we developed a model of privacy-based attacks that captures the information flows identified in each incident.
Our model defines roles for the entities involved and (in)actions that each entity performs.
The model captures the \numincidents privacy incidents from our data collection, and provides a unified view that enables further analysis.

We first identified components corresponding to the model, such as who was the subject of the data, and then used descriptive coding to assign codewords corresponding to these components of the model with respect to the article data. If something in the article did not fit with any of the model components, it was discussed within the team as part of our iterative process for deriving the final model. 
We then visualized the data flows (\Cref{sec:data-vis}) and analyzed patterns that arose,
the results of which are discussed in \Cref{sec:results-incidents}.

\ifacmversion
\else
\parabf{Analysis of Interventions.}
\fi
To understand which incidents can(not) be prevented by PETs, as well as why PETs were not deployed when appropriate, we conducted a subsequent analysis of the incidents.
For each incident described in our set of data flows, we determined the intended purpose(s) of the entities interacting with the system.
With an understanding of the system, the relevant entities' purpose(s), and the resulting harm(s), we determined whether interventions exist that would allow the purposes to be achieved while mitigating harm (\Cref{sec:system-interventions}).

\section{Article Collection and Qualitative Analysis}
\label{sec:collection}
We review our process of data collection, developing our model, and the subsequent qualitative analysis of the dataset.

\subsection{Article Collection}
If the article's contents met all inclusion criteria and no exclusion criteria, the article was included in our set.
Any ambiguous instance was assessed on a case-by-case basis through group consensus.
\ifacmversion
\else
We defer details of the process to \Cref{sec:methods}.
\fi

\ifacmversion
\else
\parabf{Sources.}
\fi
For each of the three regions we analyzed (US, UK, AU), 
we collected data from two new sources:
The New York Times (US), WIRED (US), The Times (UK), The Register (UK), The Sydney Morning Herald (AU), and ABC (Australian Broadcasting Corporation, AU).

\ifacmversion
\else
\parabf{Search Criteria.}
\fi
We collected articles using each news source's website search functionality 
with the keywords ``privacy'' and ``surveillance.''
Data collection was performed on articles that were published between 1 January 2024 and 31 December 2024; which gave us a full 365 days based on a stopping point before the time of data collection.


\ifacmversion
\else
\parabf{Exclusion Criteria.}
\fi
We used a set of exclusion criteria to exclude articles that did not contain relevant privacy incidents.
First, we 
excluded articles that were not standard text-based news (e.g., podcasts, videos, interviews, first-person narratives, and opinion articles).
We then filtered by title to exclude irrelevant topics 
(e.g., security incidents, product promotions, real estate, and tutorials).
Then, we applied exclusion criteria to the article's contents.


\ifacmversion
\else
\parabf{Inclusion Criteria.}
\fi
We required \emph{all} of the following inclusion criteria to be satisfied by the article.
\begin{enumerate}[nolistsep,leftmargin=13pt,itemsep=0pt]
    \item[\textbf{I1}] \textbf{Privacy Incident:} The article mentions a privacy violation (Def.~\ref{def:privviol}). 
    For our purposes, neither hypothetical incidents nor discussion of policies that could permit future incidents were an indication that the incidents did occur.
    Testimony that an incident occurred was a sufficient indication that it did (unless other context strongly indicated otherwise).
    \item[\textbf{I2}] \textbf{Indication of Unaccepted Practice:} The article indicates that the incident is outside the scope of accepted practices.
    This criteria excludes a myriad of out-of-scope practices\ifacmversion\else, \fi e.g., social media posts resulting in harassment, and surveillance footage leading to arrests.
    Indications that a practice is unaccepted are limited to:
    \begin{itemize}[leftmargin=7pt,itemsep=0pt\ifacmversion\else,nolistsep\fi]
        \item[-] The article includes quotes or sentiments from individuals or organizations concerned about the practice. Journalists include expert opinions to support a narrative that a given practice is (un)accepted; leveraging this allows us to respect the perceptions of incidents in reporting regions.
        \item[-] The article mentions that the practice is illegal where the event took place or where the article was published, or that there is a push to make the practice illegal/restricted.
        \item[-] The article mentions a privacy-related accusation, lawsuit, or legal case regarding the practice.
        \item[-] The practice refers to the creation or spread of non-consensual explicit imagery, which we understand to be unaccepted. 
    \end{itemize}
    \item[\textbf{I3}] \textbf{Harm:} The article mentions some form of harm that pertains to the privacy incident.
    Generally, we required the harm to be explicitly stated in the article\ifacmversion\else. \fi
    However, we identified a few practices that are understood to cause harm, but whose harms were often not concretely described.
    For these cases, we did not require elaboration: 
    \begin{itemize}[leftmargin=7pt,itemsep=0pt\ifacmversion\else,nolistsep\fi]
        \item[-] Non-consensual explicit imagery (NCEI): Users generally perceive non-consensual sharing and AI-generation of sexual images as harmful~\cite{SOUPS:BWKR24,CHI:UHBB24,CHI:SBDDG23}.
        NCEI is a subcategory of image-based sexual abuse, which 
        is widely understood to cause psychological and social harms~\cite{SLS:MJRHGFP21,IRV:Huber23}.
        \item[-] Targeted advertising: Negative user perceptions of targeted ads
        are common \cite{WPES:MC10,TKHBH09,SOUPS:ULCSW12},
        following evidence of psychological harm
        \ifacmversion
\else
        \cite{CSCW:GOS22,SP:MSBBMMA24}.
        \fi
        Equally disconcerting is the impact of politically-motivated targeted ads in elections~\cite{CJC:BG21,EJPE:LRRCC25} \ifacmversion\else and discriminatory effects~\cite{CSCW:ASBKMR19}. \fi
        Targeted ads restrict users' opportunities for choice and exercises control over their actions \cite{FAccT:WBED23, IJCLP:PCCCDM21}.
        We thus categorize targeted ads as an autonomy harm. 
        \item[-] Mass surveillance: It is well known that mass surveillance leads to chilling effects which inhibit participatory democracy~\cite{BT:Penney16,NMS:SLXW19,BDS:SFMHS23}.
        Beyond violating individual autonomy~\cite{Maras12}, mass surveillance restricts collective interests in self-determination \cite{EIT:Stahl16,Parsons15}.
        These harms are difficult to observe as they often result in lack of action.
        Thus, we consider mass surveillance a harm in its own right.
    \end{itemize}
    \item[\textbf{I4}] \textbf{Technology:} The privacy incident and/or harm must involve digital technology.
\end{enumerate}


\ifacmversion
\else
\parabf{Cross-validation.}
\fi
Approximately $1/12$ of the articles returned from initial search\footnote{For ABC, the number of articles was small, so all articles were rechecked.} 
were rechecked 
by another team member. 
Results of cross-validation are reported in \Cref{tab:cv-results}. 
Disagreements primarily arose from whether it was clear that the privacy violation directly caused the harm.
These were resolved via group discussion, and any further understanding of criteria was documented and applied to the dataset.

\begin{table}[t]
\caption{Results of cross-validation, including the total number of articles to which we applied the data collection criteria, the size of the cross-validation sample, and the percentage of articles in the sample for which there was agreement.}
\resizebox{\columnwidth}{!}{%
\begin{tabular}{lccc}
\toprule
News Source & Articles & Sample & Agreement\\
\midrule
The New York Times & 1858 & 155 & 99.35\% \\
WIRED & $>$358 & 101 & 94.06\% \\
The Times & 978 & 125 & 96.00\% \\
The Register & 649 & 56 & 96.43\% \\
The Sydney Morning Herald & 941 & 72 & 91.67\% \\
ABC & $\geq$137 & 137 & 92.70\%\\
\bottomrule
\label{tab:cv-results}
\end{tabular}
}
\end{table}

\ifacmversion
\else
\parabf{Summary of Data Collection.}
\fi
The results of data collection are summarized in \Cref{fig:numofarticlescoll}, with more details in \Cref{sec:dc-counts}.

\ifacmversion
\else
\parabf{Limitations.}
\fi
Our data collection is restricted to English-speaking regions so as not to rely on machine translation. 
Thus, our results are primarily concerned with WEIRD societies~\cite{weird}.
We leave extending the study to cover more diverse populations as interesting future work.

Due to the necessary scoping required to achieve a manageable dataset, we exclude some classes of scenarios, even though the harms that arise from them may be mitigated by privacy technologies.
For example, we excluded data breaches, distribution of dis/misinformation, and Internet shutdowns.
Our data may be biased towards events that occurred during the publication period of the articles,
or towards events that are particularly \emph{newsworthy}, since we only captured incidents that appeared in our sample of public reporting.
Still, we note that only 21.2\% of attacks in our data involve notable people or people performing noticeable actions (defined in \Cref{app:codebook}).


\begin{table}[th]
\caption{Number of articles collected and remaining after each data collection step. News sources are: NYT (The New York Times), WIR (WIRED), TIM (The Times), REG (The Register), SMH (The Sydney Morning Herald).}

\vspace{-8pt}
\resizebox{\columnwidth}{!}{%
\begin{tabular}{lccccccc}
\toprule
& NYT & WIR & TIM & REG & SMH & ABC & Total\\


\midrule
Returned by Search & 1858 & - & 978 & 649 & 941 & - & - \\
\midrule

After Format Check & 1334 & - & 835 & 619 & 824 & - & - \\

After Title Check & 1199 & 358 & - & 468 & 705 & 117 & -\\

After Contents Check & 84 & 41 & 27 & 21 & 24 & 15 & 212 \\
\midrule

Excl.~During Coding & 13 & 12 & 4 & 1 & 6 & 0 & 36 \\
Incl.~During Coding & 0 & 0 & 1 & 1 & 0 & 0 & 2 \\
\midrule

Final Count & 71 & 29 & 24 & 21 & 18 & 15 & 178 \\

\bottomrule
\end{tabular}
}
\label{fig:numofarticlescoll}
\end{table}

\subsection{Model Development and Article Analysis}
\label{sec:model-dev}



\ifacmversion
\else
\parabf{Model Development.}
\fi
To develop our model, we began with the model from Csomor~\cite{Csomor23} which observes that  
each privacy incident consists of several core components (data subject, privacy violation, attacker, and harm),
and defines the roles of initial sharer, initial receiver, and data handlers.
Over an iterative process, we revised the definitions of each role and action, and then described a set of privacy incidents. 
Our iterations concluded when we converged on a model that captured all types of data flows in our dataset.
Concurrently with our iteration process we developed a list of contextual properties that describe the conditions under which the information flows occur. 
Collectively, these two components (role--action model and contextual information) make up our model of privacy-based attacks, described in detail in \Cref{sec:model}.

\ifacmversion
\else
\parabf{Qualitative Analysis.}
\fi
We used a \emph{hybrid} approach for the descriptive qualitative analysis of our data~\cite{Saldana}.
We began with a set of predetermined codes based on the roles/actions and contextual categories in our model.
Then, we extended this set inductively based on themes that emerged from further analysis.
The resulting codes were then grouped into higher-level categories.
Our complete codebook with descriptions of each code is given in \Cref{app:codebook}.

A first pass of qualitative coding was performed by two teams of three researchers, each responsible for a subset of news sources.
Coding groups regularly discussed and readjusted themes by comparing coded extracts of articles to ensure consensus throughout the analysis.
Ambiguous cases were assessed by the entire analysis team.
Following the first pass, a second pass was performed to ensure that any changes made to coding criteria during the first pass were applied to all articles.
Finally, a review of all data was performed by a pair of researchers to ensure consistency and confirm consensus between the pair.\footnote{Measures such as inter-rater reliability are therefore incompatible with our consensus-based approach for this portion of our analysis. Rather than focusing on alignment among independent coders to verify correctness, our method focuses on mutual conceptual understanding of the codebook and subsequent documentation.}
All coding was completed manually.

\ifacmversion
\else
\parabf{Applicability of Model and Codes.}
\fi
The overall structure (roles and actions) of our model stabilized after iterating over only 20 articles (i.e., the structure and definitions of roles and actions did not change).
The stabilized model was then sufficiently generic to be applied to the remaining 158 articles.
Thus, we anticipate that the model would directly apply to other data sources, such as news articles from different years, or reports of privacy incidents from legal documents.
We expect minimal variation in codes for different articles over the same time period.
Codes might differ if the process were applied to sources from different time periods where unique attacks occur (e.g., new technologies would result in new codes); however, the structure would remain the same.

\section{Our Model of Privacy-Based Attacks}
\label{sec:model}
\ifacmversion\else\label{sec:role-action}\label{sec:context}\fi
We now present our model of privacy-based attacks, which consists of a role--action model and contextual information.
Further details on use of our model 
are given in \Cref{sec:code-details}.
In the remainder of the paper, we refer to specific articles as examples,
details of which are given in \Cref{sec:article-refs}, \Cref{tab:article-refs}.


We say that a \emph{privacy incident} is an instance of a privacy violation in our dataset (by inclusion criteria, it must have resulted in at least one harm).
A \emph{privacy-based attack} refers to the complete data flow, including the privacy violation and a resulting harm.
Each article may contain multiple incidents and each incident may result in multiple attacks (i.e., if it leads to several harms).
In the \numarticles included articles, we found \numincidents incidents
and \numattacks privacy-based attacks.


\ifacmversion
\else
\parabf{Role--Action Model.}
\fi
We develop a model with distinct roles and intermediate (in)actions that specify each privacy-based attack. 
The vital components are the data subject, privacy violation, attacker, and harm.
The remaining components describe how data flows from the data subject to the attacker.

The roles are defined as follows.
The \emph{data subject} is the entity whose personal data is in circulation.
The \emph{initial sharer} is the first entity sharing the data in circulation, and
the \emph{initial receiver} is the first entity who receives the data in circulation. 
A \emph{data handler} is one or multiple entities who receive and potentially pass on data.
Additionally, we keep track of which of the prior entities is the \emph{initiator of the privacy violation}, 
and which is the \emph{attacker} (the entity who uses the data in circulation to cause harm to the data subject).
The attacker is always the last entity in the data flow.

Each entity is connected by a particular action that describes how data flows from one to the next.
An action is a \emph{data access} if the receiving entity takes action to obtain the data. 
An action is a \emph{data sharing} if the giving entity takes action to provide the data.
When there is no clear initiator of a data transfer, or both entities are equally responsible for initiating the action, this is a \emph{data transfer}.

\begin{example}\label{ex:nyt014-40}
Consider an incident in which a mobile application offers feedback about driving behavior according to data collected while driving [Tab.~\ref{tab:article-refs}, Rows~\ref{r1} and \ref{r2}].
The app collects data beyond what the user expects and sends this data to a data broker, who sells the data to the user's insurance company.
Based on this information, the insurance company increases the driver's premiums if their driving behavior is classified as overly aggressive.

The data subject (DS) is the user driving and using the app.
Since the DS is not aware of the data collection, we say that the initial sharer and initial receiver (IS/IR) is the app and the action between DS and IS/IR is a \emph{data access}.
We consider selling/buying data to be an action actively taken by both parties, so the actions between app and data broker and between data broker and insurance company are \emph{data transfers}.
The app, data broker, and insurance company are all data handlers, and the insurance company is also the attacker since they use the data to increase insurance premiums for the DS.
The \emph{initiator of the privacy violation} is both the app and the data broker, because the selling of the data was the event identified as ``unaccepted'' in the relevant article.
\Cref{fig:model-driving} illustrates this data flow. 

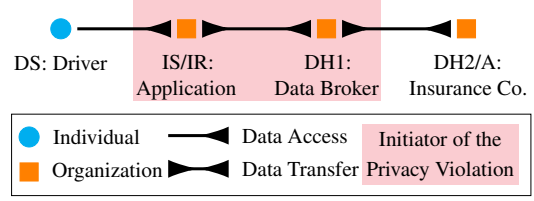
\begin{figure}[t]
\centering
\scalebox{0.8}{%
    \begin{tikzpicture}[>=Latex]
        \def\hdist{0.5}
        \def\vdist{0.5}
        \tikzset{label distance=1.5mm}

        \node[draw=none, fill=tudred!20, rectangle, fit={(1.2, -1.2) (5.3, 0.5)}, inner sep=0mm] {};

        \node [person, label=below:{DS: Driver}] (DS) {};
        \node [orga, right=\hdist*3.5 of DS, label=below:{\parbox{2.5cm}{\centering IS/IR: \\Application}}] (IR) {};
        \node [orga, right=\hdist*4 of IR, label=below:{\parbox{2.5cm}{\centering DH1: \\Data Broker}}] (DH) {};
        \node [orga, right=\hdist*4 of DH, label=below:{\parbox{2.5cm}{\centering DH2/A: \\Insurance Co.}}] (A) {};

        \draw [-{<[length=5mm,width=3mm]}, line width=0.5mm,shorten >=1mm, shorten <=1mm] (DS.east) -- (IR.west);
        \draw[{>[length=5mm,width=3mm]}-{<[length=5mm,width=3mm]},line width=0.5mm,shorten >=1mm, shorten <=1mm] (IR) -- (DH);
        \draw[{>[length=5mm,width=3mm]}-{<[length=5mm,width=3mm]},line width=0.5mm,shorten >=1mm, shorten <=1mm] (DH) -- (A);

        \node[rectangle, draw, align=left, below=\vdist*2.5 of IR, xshift=14mm] (legend) {
            \begin{tikzpicture}[>=Latex]
                \tikzset{label distance=1mm}

                \node [person, label=right:{Individual}] (indiv) {};

                \node [orga, below=\vdist*0.5 of indiv, label=right:{Organization}] (orga) {};

                \draw[-{<[length=5mm,width=3mm]}, line width=0.5mm] (2.3, -0.15) -- (3.3, -0.15);
                \node [right=2mm] at (3.2, -0.15) {Data Access};

                \draw[{>[length=5mm,width=3mm]}-{<[length=5mm,width=3mm]}, line width=0.5mm] (2.3, -0.7) -- (3.3, -0.7);
                \node [right=2mm] at (3.2, -0.7) {Data Transfer};

                \node[fill=tudred!20, rectangle, inner sep=2pt,align=center] at (6.8, 0.05) 
{\vphantom{y}\strut Initiator of the \\[2pt] Privacy Violation};
            \end{tikzpicture}
        };
    \end{tikzpicture}
    }
    \caption[Example data flow]{Example of our data flow role--action model.} 
    \label{fig:model-driving}
\end{figure}
\end{example}

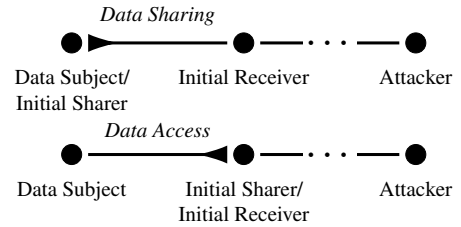
\begin{figure}[th]
\centering
\scalebox{0.8}{%
    \begin{tikzpicture}[>=Latex]
        \def\hdist{0.5}
        \def\vdist{0.5}
        \tikzset{label distance=1.5mm}

        \node [indiv,label=below:{\parbox{2.5cm}{\centering Data Subject/\\Initial Sharer}}] (DS) {};
        \node [indiv, right=\hdist*5 of DS, label=below:Initial Receiver] (IR) {};
        \node [indiv, right=\hdist*5 of IR, label=below:Attacker] (A) {};

        \draw [{>[length=5mm,width=3mm]}-, line width=0.5mm,shorten >=1mm, shorten <=1mm] (DS.east) -- (IR.west) node[midway, above,yshift=1.5mm] {\emph{Data Sharing}};
        \draw[-,line width=0.5mm,shorten >=1mm, shorten <=1mm] (IR) -- (A)
            node[midway, fill=white, inner sep=1pt] {\scalebox{2}{$\dots$}};

        \node [indiv,below=\vdist*3 of DS, label=below:{Data Subject}] (DS2) {};
        \node [indiv, right=\hdist*5 of DS2, label=below:{\parbox{2.5cm}{\centering Initial Sharer/\\Initial Receiver}}] (IR2) {};
        \node [indiv, right=\hdist*5 of IR2, label=below:Attacker] (A2) {};

        \draw [-{<[length=5mm,width=3mm]}, line width=0.5mm,shorten >=1mm, shorten <=1mm] (DS2.east) -- (IR2.west) node[midway, above,yshift=1.5mm] {\emph{Data Access}};
        \draw[-,line width=0.5mm,shorten >=1mm, shorten <=1mm] (IR2) -- (A2)
            node[midway, fill=white, inner sep=1pt] {\scalebox{2}{$\dots$}};
    \end{tikzpicture}
    }
    \caption[Possible data flows]{Possible data flows in the role--action model.
    }
    \label{fig:two-flows}
\end{figure}

The initial sharer and initial receiver roles exist to better describe the first transfer of data.
When the data subject is involved in the transfer of data (e.g., they voluntarily share data with a service), then the data subject is also the initial sharer.
Otherwise, when the data is being \emph{accessed} from the data subject (e.g., they are unaware that data is collected), then we say that the initial sharer is the entity accessing data, and the initial receiver is also the initial sharer.
In any case, these three parties are collapsed into two,
illustrated in \Cref{fig:two-flows}.

\ifacmversion
\else
\parabf{Contextual Information.}
\fi
In addition to categorizing the main entities involved in privacy-based attacks and the (in)actions that relate them, our model captures additional context surrounding the privacy incident.
This information includes the \emph{data type} (e.g., identification, location, technical, biometric, user text, demographic), \emph{entity types} (e.g., individual, government, organization), and relevant \emph{technologies} (e.g., AI, communications, news media, image/video capture, spyware).

We also record the data subject's \emph{consent} in the first transfer of data:
\begin{enuminline} 
\item unknowing, if the data subject is not aware of the data access;
\item uncontrolled, if no consent mechanism exists;
\item required for service, if the data is required by a service; 
\item under pressure/coercion, if the data subject allows sharing under some form of pressure; or
\item voluntary, if the data subject allows sharing without pressure.
\end{enuminline}

The contextual information also includes the attacker \emph{power}, i.e., amount of resources required for the attack.
Specifically, an attack with \emph{high} power requires the resources (i.e., power, money) of a government actor or
significant corporate entity, one with \emph{medium} power requires resources exceeding those of the entity being attacked, and one with \emph{low} power requires the resources of an average individual.
We also capture to what degree the attack is \emph{targeted} toward the data subject and the \emph{type of privacy violation} (i.e., collection, disclosure, unspecified transfer, or use of data).

Lastly, we classify the type of \emph{harm} brought upon the data subject.
These can be categorized according to Citron and Solove's taxonomy of privacy harms \cite{harms}: physical, psychological, economic, discrimination, autonomy, and reputational.
We found that relationship harms were captured by one of psychological or reputational.
Beyond these, we include the following categories that we understand to cause harm: NCEI and mass surveillance.

\section{Results of Analysis of Privacy Incidents}
\label{sec:results-incidents}
The following results are for our analysis of how real-world privacy incidents operate, the types of scenarios that led to each harm, and the patterns that arise in the data flow configurations.
A visualization of each data flow can be found in \Cref{sec:data-vis}.
Recall that we classify identified harms as any of: psychological, physical, mass surveillance, reputational, discrimination, economic, NCEI, and autonomy. We include these high level harms and those that fall into clusters beneath them as \Cref{tbl:highlevelharmslist} in the appendix. 
For interest, we report on the occurrences of each harm and associated codes in \Cref{sec:heatmap}.
However, we emphasize that these numbers should not be interpreted as representative of the rate of these harms nor other quantitative insights beyond what was found in our sample.

\subsection{Attacks and Harms in Privacy Incidents}

Within our sample, our model encompasses a wide range of scenarios while still revealing clear, qualitative patterns.
The harms that we identified were not limited by the technologies, levels of attacker power, or types of attackers.
Rather, we observed that each harm arises from a breadth of attacks and attack configurations. 
For this section, we have organized our harm analysis results with respect to both the level of power required to execute each attack and the degree to which attacks are targeted.
While we coded more information for each attack (e.g., legal penalties, degree of intent) to allow for analysis beyond our own, our focus here is on the harms.

\ifacmversion
\else
\parabf{NCEI, Psychological, and Physical Harms.}
\fi
Notably, within our sample,
attacks causing NCEI, psychological, and physical harms
require no particular sophistication or significant resources from the attacker (`Low' power) but, counterintuitively, are difficult to prevent through technological or regulatory means. 
Prior work observed similar technologically unsophisticated attacks in the context of intimate partner violence, finding that these attacks undermine prevalent threat models and are challenging for conventional countermeasures to address~\cite{CHI:FPMLRD16}.
Our results support this finding and emphasize the potential limitations associated with technology-first efforts towards preventing these harms.

NCEI, psychological, and physical harms are also associated with `Targeted' scenarios, where the data subject is selected as the target of the attack. 
The NCEI harms included both creating NCEI with hidden cameras and tricking a data subject into sending explicit images which are then used for extortion (termed \emph{sextortion}).
A recurring pattern regarding NCEI harms is that they involved AI and an online platform (e.g., social media).
This pattern emerges from the use of \emph{nudification} apps~\cite{USENIX:GOBCBT25}: an individual uploads an image of another person to an app which creates synthetic NCEI depicting the data subject.
Psychological harms included many of the same NCEI attacks, online harassment, and the use of GPS trackers for stalking.
Examples of physical harms that occurred in our sample are intimate partner violence and youth suicides resulting from sextortion.


\ifacmversion
\else
\parabf{Economic Harms.}
\fi
Economic harms were associated with a range of Targeting codes and 
`Medium' power attacks, meaning they are generally the next most accessible to attackers, after NCEI, psychological, and physical harms.
Economic harms we found 
involved loss of employment (e.g. an employee fired due a workplace surveillance program), heightened insurance costs (e.g. personal information influencing premiums), and fraud/extortion (e.g. phone/Internet scams).




\ifacmversion
\else
\parabf{Reputational and Autonomy Harms.}
\fi
Following economic harms, we see that reputational and autonomy harms are associated with `High' or `Medium' power.
Attacks causing reputational harms in our sample were largely `Targeted'.
This pattern may be due to the relationship between NCEI and reputational harms: the spread of NCEI itself often causes reputational harm.
Otherwise, reputational harms are often done to specific, targeted individuals.
Those in our sample included a data subject's personal information being shared publicly to shame or humiliate them, or legal prosecution of the data subject. 
Legal incidents we captured involved charges laid on the basis of facial recognition matches, DNA matches, and other technological means of investigation.

Autonomy harms we observed 
were diverse in nature: these included cases where privacy issues dissuaded patients from seeking healthcare, domestic abuse, and where a data subject's physical location was controlled (e.g. via tracking or surveillance of an area).
Due to our normative classification, targeted advertising is also included as an autonomy harm, and is perhaps overrepesented among our autonomy harms, in addition to generally being an `Untargeted' case.



\ifacmversion
\else
\parabf{Targeted Ads and Mass Surveillance.}
\fi
The attacks resulting in targeted ads and mass surveillance generally required `High' or `Medium' power and 
were associated with `Untargeted' attacks (i.e., no specific target is selected).
Like mass surveillance, targeted advertising requires access to large numbers of people and 
amounts of data~\cite{Pridmore11}.
The application of these attacks to broad publics is reflected in that they are generally performed in an `Untargeted' manner and demand significant resources from the attacker. 
The similarities between these two harms within our corpus are consistent with an understanding of targeted advertising as necessitating a form of broad surveillance.

\ifacmversion
\else
\parabf{Discrimination Harms.}
\fi
Discrimination harms surface a different type of targeting: we coded all discrimination attacks as targeted towards a `Subgroup', i.e., data is filtered to select a subset of people with a shared attribute as the target of the attack.
Nine attacks causing discrimination harms appeared in our dataset, associated mostly with `High' power. We observed no `Low' power discrimination attacks.
This likely underrepresents the number of discrimination-related cases that we encountered,
because we 
only identified a discrimination harm when the article directly mentioned discrimination. 
For example, we did not automatically code attacks involving the use of AI nudification platforms to create NCEI as discrimination harms.
However, a recent study found that most nudification platforms explicitly targeted women \cite{USENIX:GOBCBT25}, which may be considered as exacerbating discrimination.
Several attacks that involved discrimination harms had the `AI' code.
This is consistent with prior associations between AI systems and stereotyping, discrimination, and inequality~\cite{AIES:SRHMRNY23}.
The form of AI involved in these incidents was varied, including facial recognition software, algorithmic work management programs, and fraud detection algorithms.

\subsection{Our Derived Data Flow Patterns}
\ifacmversion\else\label{sec:consent}\fi
We observed an assortment of ways in which privacy-based attacks occur, demonstrated through the different patterns that arise in our data flow visualizations (\Cref{sec:data-vis}).
This assortment of patterns is a reflection of our model's ability to capture a broad range of attacks.
Recall that our model aims to amend gaps in prior frameworks that only consider a single threat actor or exclude the harms resulting from privacy violations (\Cref{sec:model-dev}).
Here, we discuss information flow patterns that can be identified as a result of the broader scope of our model.


\ifacmversion
\else
\parabf{Complex Data Flows.}
\fi
Privacy-based attacks within our corpus involve \emph{multiple} data transfers between the data subject and the attacker, where one of these transfers corresponds to the privacy violation.
We observe data flows involving up to six data transfers, a single privacy violation resulting in up to five harms, and up to three privacy violations combining to lead to the same harm.



Our data includes all types of configurations for the initial data transfer from the data subject (recall from \Cref{fig:two-flows}, the data subject may be the sharer or data may be accessed from them). 
We find that, regardless of whether or not (and to what extent) consent is provided by the data subject, nearly all harms still materialize.
Moreover, in some scenarios, the initial transfer is neither an access nor a sharing of data, but a \emph{preexisting} relation.
This occurs when an entity has prior access to an individual's data, precluding any opportunity to provide consent.
For example, in \emph{proximity advertising} [Tab.~\ref{tab:article-refs}, Rows~\ref{r16}--\ref{r17}], an individual, who shares attributes with the data subject, searches for a topic online, and the data subject receives targeted ads for the same topic because they use the same Wi-Fi network (despite not having searched themselves).
It also occurs in cases where
an individual uploads their data to an ancestry service, resulting in the prosecution of the data subject with related DNA [Tab.~\ref{tab:article-refs}, Row~\ref{r19}].

\ifacmversion
\else
\parabf{Additional Actors.}
\fi
Our model captures threat actors beyond that which ultimately causes the harm.
Notably, the attacker is not always the entity who selects the data subject as the target.
For such scenarios where the data subject is selected by an earlier party through whom the data is passed, we coded the targeting as `Targeted - Indirect'.
Such cases include the use of nudification platforms.
Moreover, the attacker may not intend to cause harm, but rather is providing a technology or service used by another entity to cause harm; here we coded the motivation of the attack as `Collateral'.
This included use of social media platforms for defamation or harassment.
Example~\ref{ex:ubereats} illustrates a `Targeted - Indirect' attack, where a service is used to cause harm and where the motivation of the attacker (UberEats) is `Collateral'.


\begin{example}\label{ex:ubereats}
    A Call of Duty (CoD) gamer (DH1) obtains the IP address of a teenager (DS) from the game [Tab.~\ref{tab:article-refs}, Row~\ref{r8}].
    They use this information to find his physical address\footnote{The article does not describe \emph{how} the individual obtained the physical address, and therefore this is not included in our data.
   In theory, there is an additional path in the data flow for how DH1 obtains the physical address.} and post this information to the gaming platform lobby (privacy violation, sharing from DH1 to DH2). 
    This led to some individuals (DH3/Attacker) harassing him and having repeated Uber Eats (DH4/Attacker) deliveries made to his house, requiring him to pay.
    \Cref{fig:ABC003} illustrates the data flow.
\end{example}

    \begin{figure}[ht]
    \centering
    \scalebox{0.75}{%
    \begin{tikzpicture}[>=Latex]
        \def\hdist{0.5}
        \def\vdist{0.5}
        \tikzset{label distance=1.5mm}

        \node[draw=none, fill=tudred!20, rectangle, fit={(1.25, -1.2) (2.45, 0.5)}, inner sep=0mm] {};

        \node [person, label=below:{DS: Gamer 1}] (DS) {};
        \node [person, right=\hdist*3 of DS, label=below:{\parbox{2.5cm}{\centering DH1: \\Gamer 2}}] (DH1) {};
        
        \draw[-,line width=0.5mm,shorten >=1mm, shorten <=1mm] (DS) -- (DH1)
            node[midway, fill=white, inner sep=1pt] {\scalebox{1.5}{$\dots$}};
            
        \node [orga, right=\hdist*3 of DH1, label=below:{\parbox{2.5cm}{\centering DH2: \\CoD Platform}}] (DH2) {};
        
        \node [person, right=\hdist*3 of DH2, label=below:{\parbox{2.5cm}{\centering DH3/A: \\Gamer 3}}] (A) {};

        \node [orga, right=\hdist*2.5 of A, yshift=-5em, label=below:{\parbox{2.5cm}{\centering DH4/A: \\UberEats}}] (A2) {};

        \draw[{>[length=5mm,width=3mm]}-,line width=0.5mm,shorten >=1mm, shorten <=1mm] (DH1) -- (DH2);
        \draw[{>[length=5mm,width=3mm]}-{<[length=5mm,width=3mm]},line width=0.5mm,shorten >=1mm, shorten <=1mm] (DH2) -- (A);

        \draw[{>[length=5mm,width=3mm]}-{<[length=5mm,width=3mm]},line width=0.5mm,shorten >=1mm, shorten <=1mm] (A)+(0,-1.1) |- (A2);

        \draw[->,thick,red] (A.north)+(0.1,0.1) to [out=50,in=90] (A2.north);
    \end{tikzpicture}
    }
        \caption[Data flow diagrams]{Example data flow diagram. \begin{tikzpicture}
    \draw[{-Computer Modern Rightarrow},line width=0.25mm,draw=red] (0,0) -- (0.5,0);
\end{tikzpicture} identifies which entity employed the attacker to target the data subject.}
        \label{fig:ABC003}
    \end{figure}

\subsection{Discussion of Incident Analysis}
Considering additional data flows and their relation to the resulting harms presents new means of understanding why certain mitigations are insufficient in preventing privacy harms.
Moreover, our modeling enables the identification of interventions from broader perspectives that otherwise could be missed (further demonstrated in \Cref{sec:sys-results}). From our results, we are also able to elucidate new nuances associated with consent and the role and responsibility of additional threat actors.

\parabf{On Consent.}
Consider the prevalence (and critique) of consent as a means for data protection.
A dominant approach to regulating data collection online is mandating ``notice-and-consent,'' where a user must be notified of and provide consent to their information being collected and used.
At the same time, there are numerous critiques of this paradigm for offering unreasonable tradeoffs, lacking sufficient information, and placing the individual in an inappropriate position to make a choice about social interests~\cite{SW14,Susser19,Nissenbaum11,Solove24,BN09,Cranor12}.

Our results support the widespread argument that consent is insufficient to prevent privacy harms in general~\cite{BarNis14}.
We identified no cases where the consent is \emph{voluntary} and the privacy violation is the first data transfer.
Additionally, under preexisting relations, there is no opportunity for the data subject to provide consent, and yet their data is distributed in a way that causes them harm.
While it is well-understood that consent is insufficient for preventing harms, our model provides a means of formally capturing why this is the case.
Despite preventing privacy violations occurring on the \emph{first} data transfer, privacy violations (and resulting harms) still occur \emph{after} this initial transfer.
More generally, as data moves away from data subjects to multiple entities via different paths, it becomes less clear how to best protect their privacy interests.

\parabf{On Responsibility.}
Our modeling does not assign responsibility to an entity for the resulting harm; rather, it identifies that each entity in the data flow plays a meaningful role.
For example, we identified cases where technologies were used by other actors to enact harm.
While some of these cases involve technologies being used for purposes for which they were designed, others demonstrate how technologies are (mis)used beyond their primary purpose.
This is consistent with prior observations in the context of intimate partner surveillance, where \emph{dual-use} technologies 
may be used for spying on a partner \cite{SP:CDOHPF18}.
This suggests that the entity ultimately causing harm may not intend to do so, and thus may be incentivized to support preventative measures (see \Cref{sec:sys-results} for more on incentives).

\section{System Intervention Analysis}
\label{sec:system-interventions}
\ifacmversion\else\label{sec:purpose-method}\label{sec:interv-methods}\fi
Having identified an abundance of privacy-based attacks, we turn to exploring by what means (technical or regulatory) could the harms be mitigated. 
From our initial analysis, we understand that there are several perspectives (both in terms of which entity and at which point in the data flow) from which one might consider harm-preventing interventions.
Thus, we employ a methodology that enables us to consider a range of perspectives 
towards identifying interventions that prevent the resulting harm. Alongside this analysis, we examine how these interventions affect the goal(s) of the entities involved.

\parabf{Categories of Interventions.}
Since entities have different capacities for enacting change, 
we consider interventions from two perspectives: users of a 
system and providers of that system.
A \emph{system provider} is the entity responsible for providing the means for users to engage with the technological system.
On their own, users do not have the means to change the system.
Their capacity to enact change is restricted to changing which technologies they use or changing how they use it.
On the other hand, system providers are in a position to redesign their system and their actions.\footnote{Analogously, in differential privacy systems, there are \emph{centralized} solutions (where a central provider applies the privacy mechanism)~\cite{EC:DKMMN06} and \emph{local} solutions (where each user applies the privacy mechanism) \cite{CCS:ErlPihKor14}.}
While we refer to regulatory interventions as \emph{self-regulatory}, these measures might be enforced by third parties. Regardless, it remains up to the entity to decide whether they abide by such restrictions.
We thus consider the following categories of system interventions:
\begin{itemize}[nolistsep,itemsep=0pt,leftmargin=0pt]
    \item [] \textbf{User-based Technical (UBT):} A \emph{technical} intervention that the \emph{user} could apply to the system in order to protect their privacy and mitigate the harm, e.g., using Tor, ad blockers.
    \item [] \textbf{User-based Self-Regulatory (UBR):} A \emph{regulatory} intervention that the \emph{user} could apply to the system in order to 
    mitigate the harm, e.g., reducing their data input into the system. 
    \item [] \textbf{System-based Self-Regulatory (SBR):} A \emph{regulatory} intervention that the \emph{system provider} could apply to the system in order to mitigate the harm, e.g., not (mis)using technologies, not (mis)using collected data, and not sharing data with third parties.
    These do not require changes to the system design.
    \item[] \textbf{System-based Privacy Engineering (SBPE):} The system could be redesigned in a way that mitigates the harm, e.g., to collect less information, offer greater transparency, ask for consent, or perform data processing locally. 
    These include strictly technical and non-technical changes.
\end{itemize}





\parabf{System Analysis Protocol.} 
Our methodology for system intervention analysis proceeds as follows.

\ifacmversion
\else
\parait{Extracting System Interactions.}
\fi
For tractability of our analysis, we focus on a particular component of each data flow. 
A data flow can be divided into components, each describing a \emph{system interaction}:
\begin{enuminline}
    \item between two entities (e.g., a financial transfer of data or someone being filmed),
    \item between a user and a service provider, or 
    \item between two entities via a service provider (e.g., posting content to an online platform).
\end{enuminline}
We analyze one system interaction for each incident.

We focus on the system interaction involving the data subject, in alignment with our interest in whether \emph{users} of technological systems can achieve their goals without incurring harms.
When the initial data transfer from the data subject is undefined, i.e., initial sharer and receiver are both unknown, then we focus on the system interaction with the privacy violation.
We call this system interaction (whether it is the one involving the data subject or the privacy violation), the \emph{primary system interaction}.
As an exception, when the initial data transfer corresponds to a \emph{preexisting relation}, the data subject is not actively using any system.
In these cases, we apply the same method, replacing the data subject with the initial sharer, i.e., the primary system interaction is the interaction involving the initial sharer, or if that is undefined, the interaction with the privacy violation.

Recall the example of driving applications 
selling data to data brokers.
The data flow consists of three system interactions: drivers using the apps, apps selling data to brokers, and brokers selling data to insurance companies.
The primary system interaction is the drivers using the applications.

\ifacmversion
\else
\parait{Enumerating Intended Purposes.}
\fi
Ultimately, our interventions should, if possible, continue to enable the primary goals of the entities involved (and if not, we wish to explore these restrictions). 
Thus, we identify the goal(s) of each relevant entity. 
For each entity involved in the primary system interaction (either user or system provider), 
we classified the entity's \emph{intended purpose} 
as one of the following:
\begin{itemize}[nolistsep,itemsep=0pt,leftmargin=10pt]
    \item[-] \textbf{None:} the entity has no purpose for interacting with the system (e.g., they are filmed unknowingly, they are the victim of an assault in which data is taken from them).
    \item[-] \textbf{Not Enough Information:} we do not have enough information to determine the entity's purpose for interacting with the system.
    \item[-] \textbf{Purpose Implies Harm:} the entity's purpose for engaging with the system 
    implies a harm (e.g., filming another during an assault, deploying facial recognition for mass surveillance).
    \item[-] \textbf{Financial Gain:} the entity's only purpose for interacting with the system is for financial gain (e.g., selling data\ifacmversion\else).\fi
    \item[-] \textbf{Purpose(s):} any purpose not in the above categories.
\end{itemize}
If 
at least one 
purpose is not classified as 
None, Not Enough Information, Purpose Implies Harm, or Financial Gain, then we include the incident in the following analysis.

For purposes that imply harm, it is contradictory to ask whether one can mitigate a harm while allowing a purpose that implies the same harm. 
This excludes some classes of incidents from further analysis, in particular, 
cases where an entity collects information for the purpose of causing harm and the data subject does not consent. 
Typical scenarios following this pattern are most cases of mass surveillance, stalking via GPS trackers or spyware, images taken during an assault, incidents of sextortion, and nudification.
Since this often occurs without the data subject's involvement or consent, they have no purpose for engaging with the system, and the individual using the application does so for the purpose of causing harm.
These are excluded, with the exception of cases where the original images were posted by the data subject on social media (here, the data subject has a valid purpose).

Scenarios where all entities only have financial gain purposes are also excluded, e.g., an exchange between a data broker and another entity who uses the data to cause harm.
These are uncommon among our primary system interactions (2/\numincidents).
If at least one entity involved in the system interaction has 
\ifacmversion
\else
an analyzable purpose,
\fi
then it is included (e.g., a user participating in a survey for financial gain and a survey provider collecting survey data for a scientific study would be included because the survey provider has another purpose).


\ifacmversion
\else
\parait{Possible Interventions.}
\fi
Finally, we determine whether there are interventions that mitigate the harm(s) while allowing the purpose(s) to be achieved.
If an entity has multiple purposes (e.g., a user engages with Instagram for entertainment and for their career), then we consider each purpose separately.

For each category of system intervention
\ifacmversion
\else
(UBT, UBR, SBR, SBPE), 
\fi
we ask whether it exists or could exist in the near future. 
If the answer was `yes' (an intervention prevents the harm), `partial' (an intervention partially mitigates or mitigates a subset of harms), or `maybe' (depending on missing details or technical developments, an intervention may mitigate the harm), then we record a descriptor of the intervention.

Here, our concept of what is considered a PET is intentionally broad.
We include as technical interventions any relevant technology that mitigates the resulting harm(s).
This scoping ensures that we do not restrict possible interventions to ``standard'' PETs 
and encompass broader techniques.
Moreover, we focus on the \emph{existence} of interventions, and not necessarily how they would be implemented.
For example, one can determine whether a PET under the umbrella of multi-party computation (e.g. private set intersection, fully homomorphic encryption) applies by asking whether a trusted third party performing the relevant computation would prevent the harm.

Upon identification of an intervention, we determined whether the introduction of the intervention \emph{restricts the purpose} of the user and/or the system provider.
For example, an intervention might impede the functionality of a system or the user's ability to freely make use of it.
These additional properties allow us to analyze whether the relevant entities have (dis)incentives regarding introducing interventions. 

For systems where an entity has a financial gain purpose, we first determine whether interventions satisfying all non-financial gain purposes exist, then in a \emph{secondary analysis} we consider interventions satisfying all purposes (incl.~financial gain).
Financial gain primarily arises as an alternative motivation of the system provider.
Separating these purposes into two analyses allows us to focus on whether the user's purpose can be achieved while preventing harm, and to identify which interventions are excluded by a requirement of financial gain.


For some cases, it is difficult to determine whether an intervention exists without domain-specific knowledge.
For example, whether one can perform molecular surveillance to track HIV clusters without sharing personally identifiable information [Tab.~\ref{tab:article-refs}, Row~\ref{r6}] is outside our domain of expertise.
In such cases, we say there is `Not Enough Information' to determine whether an intervention exists.

\begin{example}\label{ex:tiktok}
TikTok employs user data to serve targeted ads on other platforms [Tab.~\ref{tab:article-refs}, Row~\ref{r4}].
The relevant quote is:
\begin{quote}
TikTok compiled information about children using the app [...] 
then shared some of that data with Facebook [...] to help lure young users back to TikTok.
\end{quote}

The primary system interaction is between the data subject and the TikTok application.
Both the data subject and TikTok have a purpose for interacting with the system: obtaining or providing entertainment via TikTok content.
TikTok also profits financially from users viewing advertisements.
This `Financial Gain' purpose is included in a secondary analysis.


In the primary analysis, a \emph{user-based technical} intervention exists: users can use ad blockers on websites they visit.
We do not find any \emph{user-based self-regulatory} intervention, 
since the user cannot control the data collected by the app.
A \emph{system-based self-regulatory} intervention also exists and does not restrict any purpose:
TikTok need not share data with third party advertisers.
Lastly, there is a \emph{system-based privacy engineering} intervention:
the application could be redesigned to collect less data.
This intervention does, however, restrict purposes.
The user's entertainment depends on the efficacy of the recommendation algorithm, which in turn depends on the amount of user data collected.
Thus, this incurs a \emph{partial restriction} on the purpose of both user and system provider.

\end{example}

\ifacmversion
\else
\parabf{Positionality and Reliability.}
\fi
This analysis was performed independently by three team members with knowledge of and experience with a range of privacy technologies, including secure computation (e.g., fully homomorphic encryption, multi-party computation, and zero-knowledge proofs), differential privacy, privacy in machine learning, and cryptography.
We required that at least two of three team members agreed that the identified intervention was plausible/correct.
In general, this meant that we took the \emph{union} of the three 
experts insights, opting for the most positive decisions.

\section{Results of Systems Analysis}
\label{sec:sys-results}
We provide general results for system interventions and discuss the scenarios for which we identify no intervention.
For the remaining systems, we analyze disincentives to deployment 
and scenarios with no purpose restrictions. 

\subsection{Purposes Implying Harm}
\label{sec:bad-purposes}
From our set of \numincidents incidents, we excluded systems without analyzable purposes, 
leaving \numsystems systems.
Of the systems that were excluded from further analysis, the systems were almost exclusively ones where they included an entity whose purpose implied the harm. 
It is impossible in such cases to attempt to mitigate the harms with PETs while enabling the purpose, since the purpose itself is to cause harm.

This process excludes certain scenarios that may be of interest.
For example, our data includes several reports of AirTags and GPS trackers being used for stalking [Tab.~\ref{tab:article-refs}, Rows~\ref{r12}--\ref{r15}, \ref{r18}].
These were excluded from our systems analysis because in the incidents of stalking, 
the entity collecting data was only doing so to cause harm.
Yet, there is ongoing research and interest among privacy technologists about the risks of GPS trackers and potential mitigations~\cite{USENIX:GFGDK25,PoPETS:SGNSMR23,PoPETS:HeiWurHol24,USENIX:EBGHJ24}.
This also excludes most cases of mass surveillance; incidents where technology was used for assault, abuse, or harassment; technology used to deter homeless individuals from loitering; many instances of nudification applications; and scenarios involving technology-driven scams, blackmail, and extortion.

While some of these systems are primarily harmful, other technologies often enabled dual purposes, where some purposes may not be harmful.
For instance, GPS trackers can be used for tracking one's own devices, but can also be used for stalking.
Our 
analysis does not account for alternative purposes of the same technologies, e.g., finding lost devices.
Notably, PETs can be counterproductive to preventing harms when enabling certain functionalities (e.g., unlinkability and detectability of stalking are contradictory goals for GPS trackers~\cite{EPRINT:KOJRS26}).
We discuss a need for exploring interventions and implications of such systems in \Cref{sec:implies-harm}.

\subsection{Potential Interventions}
We consider the remaining \numsystems systems with analyzable purposes.
Of the \numsystems cases in our set, there are 59 where at least one intervention type mitigates the harm completely (`Yes'), and 85 where at least one intervention type may apply or partially mitigates the harm (`Yes', `Maybe', or `Partial').
The cases with no relevant intervention include:
\begin{enuminline}
    \item redistribution of publicly available content, resulting in harassment or economic harm;
    \item employee surveillance tools tracking productivity, resulting in job loss;
    \item use of a database by real estate agents to track tenants who missed rental payments, resulting in denial of housing applications; and
    \item YouTube training on content for AI services, resulting in harm to job prospects of content creators\ifacmversion\else.\fi
\end{enuminline}

Among the scenarios that we considered, system-based interventions exist more often than user-based interventions, which is consistent with system providers typically having more power to enact change.
Self-regulation, including not (mis)using collected data or not sharing data with a third party, can completely mitigate harms, while privacy engineering 
more often partially mitigates harms.
From a user's perspective, self-regulatory and technical means are similarly effective, but apply in a minority of cases that we considered.

Several technological strategies emerged as useful for mitigating harms.
First is more fine-grained \emph{access control}: either the user or the system restricting access to information can prevent unwanted parties from causing harm.
In some cases, a psychological harm arises due to the data subject's fear of what might arise from data sharing.
For these situations, it may be sufficient to provide \emph{transparency} about the use of data and to provide the user an opportunity to \emph{consent} to data collection.
Sometimes, e.g., in doxxing incidents, it is sufficient to \emph{detach identity} from the data subject's actions or other information. 
Though useful in some cases, we note that consent and anonymity are insufficient for preventing harm in general. 
Finally, \emph{data minimization} remains a useful strategy for preventing privacy violations.

More examples and further result details are provided in \Cref{sec:interventions}.
Crucially, these techniques for designing (technical) interventions are insufficient in isolation.
While the existence of a private alternative to a system may encourage adoption, one must still address misaligned incentives and changes in power dynamics in parallel to technological developments.

\subsection{(Dis)incentives to Deployment}
\label{sec:disincentives}

Given that there exists an intervention for the majority of scenarios in our dataset, we explore why such interventions were not implemented.
A significant body of research considers the challenges faced by users in adoption of PETs~\cite{PETS:RenVolRen14,USENIX:WhiTyg99,PoPETS:NWCK20,PoPETS:SSYACS21,PoPETS:HarPapRan20}; here, we focus on the restrictions for those technologies and regulatory solutions that prevent harm in the real-world scenarios surfaced in our analysis. 

\begin{figure}[t]
\centering
\includegraphics[scale=0.25]{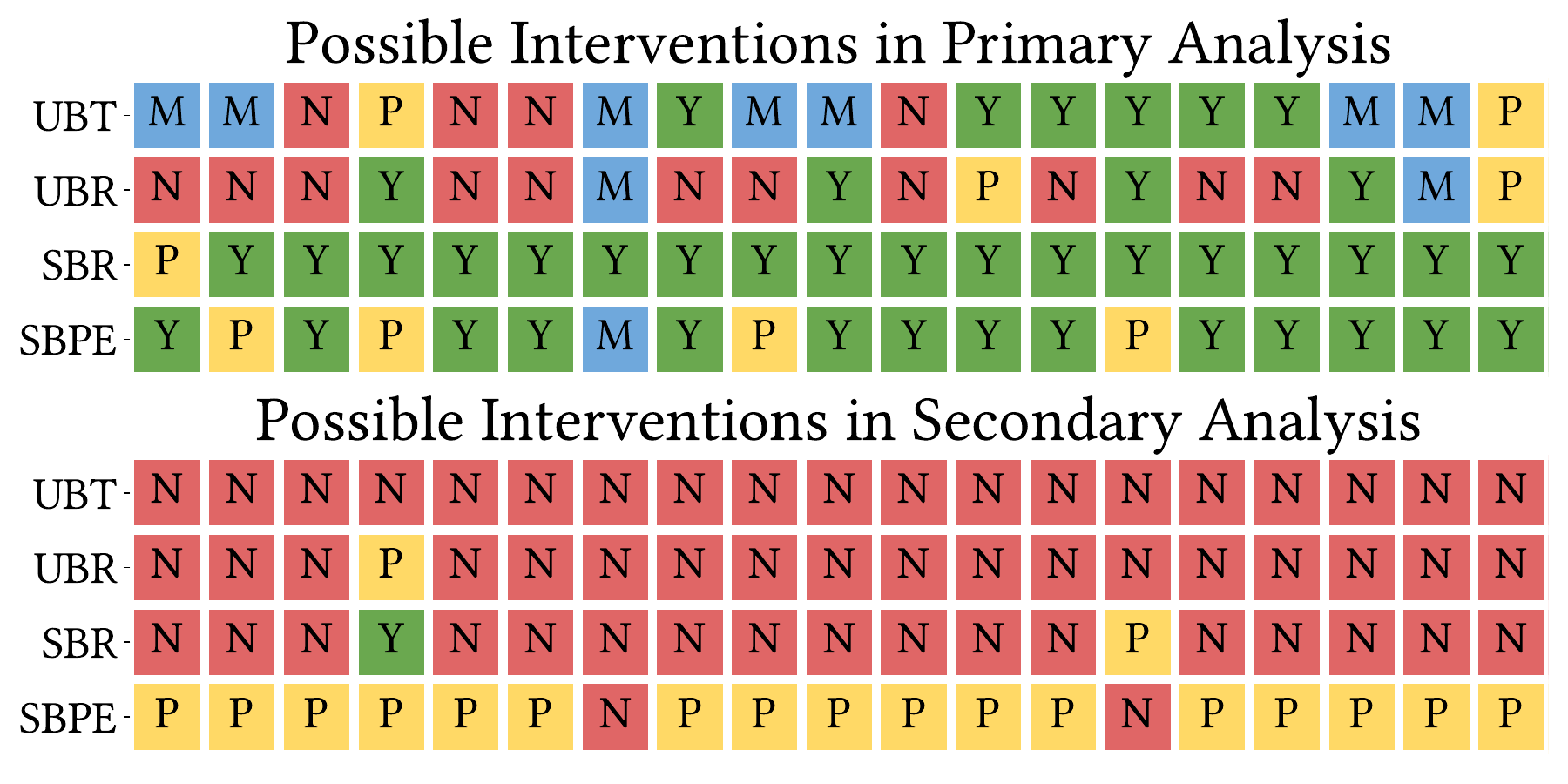}
\caption[Primary vs Secondary Analysis]{Primary vs Secondary Analysis. Each column corresponds to one analyzed system.
The top rows show which intervention types apply for the system in a primary analysis (no financial gain purpose considered), and the bottom rows show results for the same system including a financial gain purpose. 
\tikz[baseline=(square.base)]{
  \node[draw=green!60!olive!40, fill=green!60!olive!40, rectangle, inner sep=1pt] (square) {Y};
} = Yes,
\tikz[baseline=(square.base)]{
  \node[draw=red!50, fill=red!50, rectangle, inner sep=1pt] (square) {N};
}
= No,
\tikz[baseline=(square.base)]{
  \node[draw=orange!30!yellow!50, fill=orange!30!yellow!50, rectangle, inner sep=1pt] (square) {P};
}
= Partial,
\tikz[baseline=(square.base)]{
  \node[draw=cyan!45, fill=cyan!45, rectangle, inner sep=1pt] (square) {M};
}
= Maybe}
\label{fig:priv-vs-sec}
\end{figure}

\ifacmversion
\else
\parabf{Profit vs Harm Prevention.}
\fi
Preserving the system provider's ability to profit financially severely limits potential interventions.
\Cref{fig:priv-vs-sec} shows for each system with both a primary and secondary analysis (i.e., where we considered financial gain in a secondary analysis) whether each intervention type exists.
Most systems with financial incentives in our set are those where a user accesses an online service and harms arise from the service provider \begin{enuminline} \item training on user data, \item selling user data to a third party, or \item performing targeted advertising\end{enuminline}.
In the primary analysis, a user can sometimes use an ad blocker or similar technology to prevent harms, while the system can almost always be regulated or redesigned in a way that prevents the use of information that results in harm. 

Once financial gain is included as a purpose in a secondary analysis, 
the systems can often still be redesigned to share aggregate information (e.g., using differential privacy) or target advertising towards groups of users rather than individuals.\footnote{We did not recognize tools for \emph{private advertising}, which allow providers to serve targeted ads without directly revealing personal information, as relevant interventions.
These systems protect data confidentiality, but do not prevent the resulting harm~\cite{Veale23,PoPETS:HCKVD26}.}
These aggregation-based methods partially mitigate harms while often restricting the system provider's ability to profit.
Furthermore, they may be undesirable because they may exacerbate discrimination against certain groups~\cite{ULJ:Allen19}.


\ifacmversion
\else
\parabf{Purpose Restrictions.}
\fi
A clear pattern arises from our analysis of restrictions: user-based interventions generally restrict the purpose of the user 
while system-based interventions generally restrict the purpose of the system provider. 
For interest, 
\Cref{fig:restriction-counts} in \cref{sec:interventions} illustrates whether the identified interventions 
restrict the purpose of the user and/or the system provider.

Tradeoffs between privacy and utility are frequently referenced, particularly in relation to publishing data~\cite{KDD:BriShm08,KDD:LiLi09,PODS:DinNis03} and machine learning~\cite{ICML:SKGPT24,ESA:CMFA23}.
Moreover, it is well-known that the deployment of PETs may be inhibited due to lack of usability (from the perspective of users)~\cite{USENIX:WhiTyg99,SOUPS:ABHLG15}.
We observe that the introduction of privacy interventions not only decreases the utility of a system (for both users \emph{and} service providers), but also introduces additional types of restrictions.


Restrictions imposed on users fall into three categories: (1) decreased utility of the system, (2) decreased usability of the system, and (3) restricted freedom to perform tasks adjacent to the system.
Utility of the system may be affected by a reduction in functionality (e.g., not including a recovery email address for an account means losing access if a password is forgotten, but allows for greater anonymity), a reduction in freedom to use the system (e.g., posting less on social media), or a reduction in efficacy of the system (e.g., providing less information to algorithm-driven applications reduces their ability to tailor content).
Usability of the system is particularly affected when the intervention involves obfuscation techniques (e.g., obfuscated data in a fertility tracking app).
Lastly, interventions such as turning WiFi off on mobile devices, or using GPS spoofing would prevent tracking and data collection in some scenarios (e.g., on public transit or when visiting health clinics), but restrict the user's ability to freely use other applications or features.



Restrictions imposed on system providers fall into three categories: (1) reducing efficacy of surveillance and enforcement, (2) reducing efficiency of processes, and (3) limiting ability to perform research.\footnote{We excluded restrictions such as financial or labour costs of redesigning systems, as otherwise the answer to whether an intervention causes a restriction would always be yes.}
For example, private contact tracing (not allowing for strict quarantine enforcement), and restricting police usage of DNA information from genealogy sites both reduce the efficacy of surveillance.
Example~\ref{ex:schools} also illustrates restrictions on efficacy of surveillance.
Interventions may reduce the efficiency of processes by disallowing automated tools, e.g., facial recognition tools at airports, employee monitoring software, automated gunshot reporting, and image classification for insurance assessments.
Some interventions limit organizations' ability to perform research (e.g., requiring to ask participants for explicit consent to use their data).

\begin{example}\label{ex:schools}
Schools fit devices with surveillance technology to ensure safe Internet use among students [Tab.~\ref{tab:article-refs}, Row~\ref{r5}].
These tools perform real-time scanning of students' typing and alert staff members in situations of concern; but, these alerts can result in unnecessary contact with police.
To avoid potential psychological harms, one might propose a system design that blocks access to harmful websites 
rather than integrating automatic scanning and alerts.
This intervention may reduce the harm stemming from unwanted interactions with law enforcement, but also restricts the school's ability to monitor and control student behavior.
Prior work also noted these conflicting incentives among students and administrators~\cite{SOUPS:CTKNKS21}.
\end{example}


\ifacmversion
\else
\label{sec:leftovers}
\parabf{No Restrictions.}
We briefly review the systems where there exists a potential intervention and for which there are \emph{no restrictions} 
(also excluding cases with financial gain purposes).
\fi
Several of these scenarios were accidental data sharing that could be prevented with better data management practices (e.g., finer grained access control, data separation).
In the remaining scenarios, 
additional data was collected that was \emph{not necessary to achieve the desired purposes}.
These could be resolved with existing technologies that do not impede any functionality (e.g., phone privacy screen protectors, end-to-end encrypted messaging, VPNs, anonymous reporting tools),
or by straightforward system-based self-regulatory or technical interventions (e.g., Tesla employees need not have access to private car footage, a college can inform students to clean their dormitories without including photos of their rooms).

\subsection{Discussion of Intervention Analysis}
\ifacmversion\else\label{sec:implies-harm}\fi

\ifacmversion
\else
\parabf{Functionality-Enabled Harms.}
\fi
In \Cref{sec:bad-purposes}, we identified that many systems were excluded due to their purposes implying harm.
For many of these cases, the technological system enables some functionality that is \emph{not} the primary intention of the system, but nonetheless implies harm (e.g., use of technologies for stalking and abuse).
While a consideration of these systems is incompatible with our analytical framework, it suggests a need for deeper exploration of the purposes enabled by certain systems.
Future work may consider, for technological systems whose purposes imply harm, whether there exists interventions that can \emph{restrict the purposes/functionality} in some way that prevents the resulting harms, e.g., \cite{EPRINT:KVRSC26}.
Addressing these harms requires a new understanding of PETs beyond those which solely aim to enable functionalities while providing confidentiality and/or anonymity of data.

\ifacmversion
\else
\parabf{Realigning Incentives.}
\fi
In \Cref{sec:disincentives} we identified that for cases in which an intervention applies, the deployment of such an intervention would result in restrictions upon the entity with the capacity to deploy the intervention.
Barriers to deployment are not restricted to lack of awareness, existence of technological solutions, and usability issues.
System providers regularly offer services to users while having alternative motives, which may conflict with harm-preventing measures.
Prior work identifies similar incentive dynamics in the context of algorithmic optimization systems~\cite{FAT:KOTG20}.
Thus, further evaluation of system providers' reasoning for (non-)adoption of PETs is of importance~\cite{IEEE:PPOAJN25,USESEC:SHK26,CHI:ABVLS21}.
In general, misaligned incentives cannot solely be resolved by technical means.
Through the introduction of meaningful consequences for service providers that fail to prevent harms, this imbalance of incentives may be corrected in favour of data subjects~\cite{crawford2014big,PoPETS:SSCJGM24}.

Even when all incentives are aligned for the deployment of PETs, 
there remain systems lagging behind state-of-the-art privacy practices, and where privacy was not made a priority, either by the system provider or the user (\Cref{sec:leftovers}).
Such oversights unnecessarily lead to harms.
This may be attributed to lack of awareness of privacy as a concern, lack of concern for privacy, or not viewing a need to act based on privacy concerns (e.g., due to a lack of understanding of the possible consequences of not protecting privacy)~\cite{PETS:RenVolRen14,PoPETS:GerReiVol19}.


\section{Conclusion}

In this work, we performed an analysis of real-world privacy incidents to gain an understanding of their mode of operation and resulting harms.
Through modeling the complexity of data flows not captured by prior frameworks,
we were able to uncover new patterns in how harms emerge, and explanations for why interventions, like consent, are inadequate at preventing harms in general.
With this perspective, we considered the applicability of technological and regulatory means for mitigating harms.
Our results identified fundamental barriers to harm prevention in systems whose functionalities imply harm,
illustrating a significant gap in a research domain that largely focuses on \emph{enabling functionality} while providing privacy.
A new conceptualization of PETs is required to address these harms.
Our analysis also identified that the primary barrier to harm mitigation is not necessarily that technical interventions are unavailable or unusable, but that the party best positioned to deploy them is often the party least incentivized to do so.
Technological solutions alone cannot reconfigure misaligned incentives, and regulatory changes are necessary to introduce consequences for enabling harms.

\cleardoublepage
\appendix
\section*{Ethical Considerations}
No private information was used in our data; our data collection and analysis was performed only on publicly available information.
Given the occasionally sensitive and distressing nature of the topics covered in the articles, we performed regular check-ins with team members, allowed for frequent breaks from reading the material, and always enabled individuals to opt out of reading a particular article.

\section*{Open Science}
Our complete data (including data collection, article coding, and systems analysis) is available at \url{https://anonymous.4open.science/r/privacy-harms-data-CF9E/}.


\ifacmversion
\else
\bibliographystyle{plainurl}
\fi
\bibliography{local.bib}


\section{Detailed Methods of Data Collection and Qualitative Analysis}
\label{sec:methods}

\subsection{Data Collection}

Our data collection procedure was as follows.

\paragraph{Selecting Sources}
We collected data from six news sources across three regions: The New York Times (US), WIRED (US), The Times (UK), The Register (UK), The Sydney Morning Herald (AU), and ABC (AU).
To select publications we referred to the popularity of each source in its respective region, and we verified that each publication was a reasonable choice with contacts who lived in that region.

WIRED and The Register are technology-focused publications, while the other four are more general publications.
We originally aimed for one technology-focused and one more general news source for each region, but were unable to find a technology-focused news source from Australia that returned a substantial number of articles related to our keywords.
For UK news sources, we additionally tested BBC and The Guardian, but found the search UIs to be incompatible with our data collection procedure.

\paragraph{Date Range}
We collected articles published between 1 January 2024 and 31 December 2024, inclusive. This was the most recent calendar year at the time of data collection.

\paragraph{Keywords}
We collected news articles by searching each news source for particular keywords.
The keywords we started with were ``privacy'' and ``surveillance'', as well as several keywords associated with relevant privacy regulations for each region (``CCPA'', ``HIPAA'', and ``COPPA'' for the US; ``Data Protection Act'' and ``GDPR'' for the UK; ``OAIC'' for AUS).
Over the course of the data collection process we found that the law-related keywords returned very few relevant results beyond those that were already returned by the ``privacy'' and ``surveillance'' searches.
We discontinued using the law-related keywords for The New York Times, since for this source we already had a very large number of articles from the ``privacy'' and ``surveillance'' searches.

\paragraph{Search}

For The New York Times, WIRED, The Register, and ABC, we used the search function on the publication's website and the available sort/filter operations to obtain articles.
For The New York Times in particular we also used the publicly-available New York Times API\footnote{\url{https://developer.nytimes.com/}} to produce an accompanying spreadsheet of article titles, publication dates, and URLs in the interest of reducing data entry error.

For The Times: initially we used the website search function to obtain articles.
We then realized that the website was limited to displaying only the most recent 100 pages of articles, which was not enough to cover the keywords for our entire date range.
For part of the date range we used the ``Site Map'' page, which lists all content published on the website, to obtain more articles.
We then decided to obtain further articles by querying the NexisUni database with the news source, the keywords ``privacy'' and ``surveillance'', and the target date range.

For The Sydney Morning Herald: we found the website search function insufficient for our purposes. 
Initially we used Google Advanced Search with particular arguments (e.g., \texttt{site:smh.com.au "privacy" after:2024-01-01 before:2024-02-29}) to obtain as many articles as possible with the desired keywords and in the desired date range.
Later, we decided to use NexisUni to obtain further articles from this source for the keywords ``privacy'' and ``surveillance'' (as we had for The Times), and combined the articles we had already obtained with those from NexisUni.

\subsubsection{Initial Criteria}
After collecting articles, we applied a series of criteria to exclude any that were not relevant to our analysis.
Our initial criteria were based on those from \cite{Csomor23}.
We would continue to refine these criteria during the data collection process.

\paragraph{Format exclusion}
First, we checked the format of the article.
We excluded an article if it met any of a set of criteria, aiming to exclude search results that were not standard text-based news articles.
(These criteria did not meaningfully change over our data collection process---descriptions are given in the \emph{Final Criteria} section below.)

\paragraph{Title exclusion}
Then, we applied exclusion criteria to the title of the article.
An article was excluded if its title met any of a set of exclusion criteria: 
\begin{enuminline}
    \item[\textbf{E1}] Data Security Incident,
    \item[\textbf{E2}] Promotion,
    \item[\textbf{E3}] Tutorial, or
    \item[\textbf{E4}] Real Estate and Travel.
\end{enuminline}
(As above, descriptons are given in the \emph{Final Criteria} section.)

\paragraph{Contents exclusion}
After that, we looked at the contents of each article.
An article was excluded if its contents met any of a set of exclusion criteria:
\begin{enuminline}
    \item[\textbf{E1}] Data Security Incident,
    \item[\textbf{E2}] Promotion,
    \item[\textbf{E3}] Tutorial,
    \item[\textbf{E4}] Real Estate and Travel,
    \item[\textbf{E5}] Opinion Article, or
    \item[\textbf{E6}] Duplicate.
\end{enuminline}
(As above, descriptions are given in the \emph{Final Criteria} section.)

\paragraph{Contents inclusion}
Finally, if an article had not been excluded by any of the prior checks, we applied inclusion criteria to the contents of the article.
We required all of the following inclusion criteria to be met by the contents of the article (if any one was not met, we excluded the article).

\begin{enumerate}[leftmargin=15pt]
    \item[\textbf{I1}] \textbf{Privacy Violation:} The article mentions a \emph{privacy violation}, i.e. an action or inaction that causes an entity (the data subject) to lose control about how, when, or to what extent personal information about it is communicated to others, outside of the scope of accepted practices.
    \item[\textbf{I2}] \textbf{Attacker / Harm:} The article mentions one or both of an attacker (i.e. an entity who uses the data in circulation to cause harm on the data subject) or a harm (i.e. physical, mental, or another kind of damage or worsening of state of the affected party).
    \item[\textbf{I3}] \textbf{Technology:} The privacy violation must have been caused or facilitated by technology.
\end{enumerate}

\subsubsection{Initial Data Collection Process}
One researcher was initially assigned to do the data collection for each of the six news sources (later, we had multiple researchers assisting for some sources).
When an individual found the content of an article to be confusing or was uncertain whether the article met the criteria, they shared it with other team members.
Some of these cases could be resolved by a second reading from another researcher; in other cases we discussed the article as a group and made a collective decision.
As we progressed through data collection and encountered new situations we continued to adjust our criteria through group discussion.


\subsubsection{Final Criteria}
Over the course of data collection we refined our procedure.
Our final set of steps by the end of data collection is as follows.

\paragraph{Format exclusion}
First, we checked the format of the article.
We excluded an article if it met any of a set of criteria:

\begin{enumerate}
\item The article is not primarily text based (e.g., podcasts, videos).
\item The article is not published under the news source proper (e.g. articles from Wirecutter, a publication that is owned by The New York Times Company but is distinct from The New York Times).
\item The article is not a standard format article (e.g., PDF files, Q\&As, interviews, or first-person narratives).
\item The article is in the news source's ``Opinion'' category or similar.
\end{enumerate}

\paragraph{Title exclusion}
Then, we applied exclusion criteria to the title of the article.
An article was excluded if its title met any of a set of exclusion criteria: 

\begin{enumerate}[leftmargin=15pt]
    \item[\textbf{E1}] \textbf{Data Security Incident:} The article focuses on a hack, a data breach, a software bug, malware, or an instance of a technological system not functioning as intended.
    \item[\textbf{E2}] \textbf{Promotion:} The article focuses on advertising, promoting, or reviewing products, media, or experiences.
    \item[\textbf{E3}] \textbf{Tutorial:} The article focuses on making recommendations in a tutorial or guide format.
    This includes articles with titles like ``How to be safe...'', ``How to protect your privacy...'', or ``The best XY for Z...'', which generally do not discuss specific incidents.
    \item[\textbf{E4}] \textbf{Real Estate and Travel:} The article focuses on real estate, home design, or reviewing/promoting tourist destinations.
    These often referenced locations with a ``sense of privacy,'' or private pools, villas, etc., without reference to a privacy incident.
\end{enumerate}

\paragraph{Contents exclusion}
After that, we looked at the contents of each article.
An article was excluded if its contents met any of a set of exclusion criteria:

\begin{enumerate}
    \item[\textbf{E1-4}] E1, E2, E3, and E4 from before.
    \item[\textbf{E5}] \textbf{Opinion Article:} The article focuses on convincing the reader of an opinion (e.g. op-eds). 
    \item[\textbf{E6}] \textbf{Duplicates:} The article focuses on the same incident as another article. We include the article with more detail. At times, we compile information together from an included article and some of its duplicates.
\end{enumerate}

\paragraph{Contents inclusion}
Finally, if an article had not been excluded by any of the prior checks, we applied inclusion criteria to the contents of the article.
We required all of the following inclusion criteria to be met by the contents of the article (if any one was not met, the article was excluded).

\begin{enumerate}[leftmargin=15pt]
    \item[\textbf{I1}] \textbf{Privacy Incident:} The article mentions a privacy incident, i.e., an action or inaction that causes an entity to lose control about how, when, or to what extent personal information about it is communicated to others.
    For our purposes, neither hypothetical incidents nor discussion of policies that could permit future incidents are an indication that the incidents did occur.
    Testimony that an incident occurred was a sufficient indication that it did (unless other context strongly indicated otherwise).
    \item[\textbf{I2}] \textbf{Indication of Unaccepted Practice:} The article indicates that the incident is outside the scope of accepted practices.
    Such indications included:
    \begin{itemize}[leftmargin=8pt,itemsep=0pt]
        \item[-] The article includes quotes or sentiments from individuals or organizations concerned about the practice.
        \item[-] The article mentions that the practice is illegal where the event took place or where the article was published, or that there is a push to make the practice illegal/restricted.
        \item[-] The article mentions a privacy-related accusation, lawsuit, or legal case regarding the practice.
        \item[-] The practice refers to the creation or spread of non-consensual explicit imagery, which we understand to be unaccepted. 
    \end{itemize}
    \item[\textbf{I3}] \textbf{Harm:} The article mentions some form of harm that pertains to the privacy incident.
    Generally, we required the harm to be explicitly stated in the article---articles that described privacy incidents without stating a concrete harm were excluded.
    However, we identified a few practices that appeared noticeably often in our data and are understood to cause harm, but whose harms were often not concretely described by articles.
    For these cases, we did not require additional elaboration on the concrete harms of the practice.
    Such cases were:
    \begin{itemize}[leftmargin=8pt,itemsep=0pt]
        \item[-] Non-consensual explicit imagery (NCEI)
        \item[-] Targeted advertising
        \item[-] Mass surveillance
    \end{itemize}
    \item[\textbf{I4}] \textbf{Technology:} The privacy incident and/or harm must involve digital technology.
\end{enumerate}

\paragraph{Linked articles}
Occasionally, we noticed that information about an incident required by our inclusion criteria was missing from the article but present in an another article that it linked.
In these cases we sometimes combined relevant information from both articles to meet the inclusion criteria.
We only used linked articles if they were published during our target date range and by the same news source as the original article.


\subsubsection{Cross-Validation}

After the initial data collection we performed a cross-validation to evaluate consistency.
For each news source, we assigned an individual who had not helped perform the initial data collection for that source to redo data collection on a sample of articles.

For The New York Times and The Register: the sample was a random selection of 1/12th of the articles that were returned by our keyword search.

For The Times and The Sydney Morning Herald: the sample was a random selection of 1/12th of the articles that were returned by our query to NexisUni.

For ABC: relatively few total articles appeared in search results for our final keywords (``privacy'' and ``surveillance''), and so for cross-validation we used all articles that were returned by a search for these keywords.

For WIRED: due to the way the website's search function worked and the way we had recorded articles, we were unable to sample from all articles returned by the search.
We estimated the number of articles that would have appeared in the search, and then randomly sampled around 1/12 of that number from the articles that had passed the title check (these we had recorded).

The results of the cross-validation are given in \Cref{tab:cv-results}.
For any articles where the decisions made by the researcher who had done the data collection and the researcher doing the validation differed, the two researchers discussed the article to reach a consensus on whether it should be included.
We also looked over the general sources of disagreement as a group to identify areas where individuals had applied the procedure or interpreted the criteria differently.


\subsubsection{Data Collection Recheck}

We performed a second pass over the data in the interest of improving internal consistency and addressing areas of disagreement.
At least all articles that had been included were revisited.
For most news sources we also performed additional checks, e.g. a researcher revisited all articles that had been marked as passing a certain number of inclusion criteria, or that had notes marking them as uncertain decisions.

\subsection{Article Analysis}

Our data collection yielded a number of articles that contained one or more privacy-based attacks.
We initially began with the model of privacy-based attacks and a subset of the codes from \cite{Csomor23}.
We iteratively applied the model and codes to incidents in our data and revised them to account for new situations and better represent the data flows we observed.
Our final model is described in \Cref{sec:model}, and our final set of codes is given in \Cref{app:codebook}.

\paragraph{Duplicate and linked articles}
In some cases we used relevant information about an incident from multiple related articles.
These included articles that had been marked as duplicates in our data collection procedure and articles that were linked from the original article.
As we did in data collection, we only used linked articles that were published in our target date range and by the same news source as the original article.

\subsubsection{First pass}
Initially, we performed coding in two groups of three researchers, each of which coded a subset of the articles.
We added, removed, and redefined codes as we encountered situations where the existing codes did not capture some important aspect of an incident or recurring theme in the data.

When a coding group was uncertain about how to apply the model, which code to use, or whether a new code should be added, they shared the article with the group (as we did in data collection).
Some of these cases could be resolved by a second reading from another researcher or the other coding group; in other cases we discussed the article as a group and made a collective decision.

We also encountered a number of articles that we decided should not have been included under the data collection criteria, and excluded them from coding.

\subsubsection{Second pass}
Because we had modified our codes and model over the initial process, we revisited some of the incidents after the first pass.
We combined the two groups of three into a group of six and went over a subset of the articles.
Our aim was to identify areas where we had to update codes due to changes we had made over the first pass of coding, or due to disagreement or inconsistency between the two groups.
During this process we continued to discuss any cases where the group was uncertain, modify the codes to better represent the data, and exclude any remaining articles that we decided did not meet the data collection criteria.

\subsubsection{Third pass}
Finally, two researchers looked over \emph{all} of the incidents in the dataset.
Here we aimed to check for mistakes and ensure the usage of codes was consistent with the procedure we had established.
We also made several more tweaks to our codes and added/removed several articles that we found had been missed or should have been excluded.

\subsection{Model Differences from \cite{Csomor23}}
Our role--action model and contextual information are based on the model from Csomor~\cite{Csomor23}.
The original model defines five roles (data subject, initial sharer, initial receiver, data handler, and attacker) with data accesses/sharings occurring between entities.
Our model retains the same general structure, but we made some modifications during the coding process to capture the data flows we encountered. These are explained below.
Our system intervention analysis (\Cref{sec:interv-methods}--\ref{sec:sys-results}) is new.

\subsubsection{Initial Receiver, Initiator of the Privacy Violation, and Privacy Violation}
In \cite{Csomor23}, the initial receiver is defined as the ``first entity who receives the data in circulation and causes a privacy violation.''
We found that, in many cases, the first entity who received the data was not the same entity who caused the privacy violation.
Consider a case where a data subject initially uploads images to social media (the data subject and initial sharer are the same entity, and the social media platform is the initial receiver), and then another individual (a data handler) downloads those images and uploads them to a nudification application.
This final transfer is the privacy violation, but is clearly caused by the data handler and not the initial receiver.

To describe these data flows, we had to allow the privacy violation to occur on any transfer between entities, not only transfers caused by the initial receiver (i.e. between data subject and initial sharer/receiver, between initial sharer/receiver and data handler, or between two data handlers).

Then, for compatibility, we split the original definition of the initial receiver into two simpler definitions: the initial receiver (the first entity who receives the data in circulation) and the initiator of the privacy violation (the entity who causes the privacy violation).
Since we observed privacy violations occurring in almost every position in the data flow, the initiator of the privacy violation was a flexible label which at times applied to one or more of the initial sharer/receiver, data handlers, and attacker.

\subsubsection{Ambiguous Data Transfers} 
We encountered many cases where we could not clearly determine whether a data transfer was an access or a sharing, either due to a lack of information in the article, or sometimes because both parties could be said to have taken action in the transfer (e.g., we considered buying/selling data to be a case where both the buyer and seller take initiative to perform the transfer).
We thus enabled a data transfer to be considered an ambiguous transfer instead of strictly an access or sharing.

\subsubsection{Broadened Access and Sharing}
Both models share a definition of a data access (the receiving entity takes action to obtain the data) and data sharing (the giving entity takes action to provide the data).
In \cite{Csomor23}, a data access was present between data subject and initial sharer, and between data handler and attacker.
A data sharing was present between initial sharer and initial receiver.
We found this to be restrictive since data sharing often occurred later in the data flow.
For example, we saw cases where an individual (data handler) shares the data subject's photos with another service (attacker), meaning a data sharing was present between data handler and attacker.
To capture cases such as this, in our model every data transfer between entities may be a data access or sharing (except for the transfer from the initial sharer, which is still typically a sharing).

We also encountered some cases where there was no clear data sharing among the first few entities in the data flow (e.g., the initial transfers of data were accesses or ambiguous transfers) or no sharing in the data flow at all (e.g., a data subject is unknowingly filmed by an attacker, creating NCEI).
In these cases, 
the initial sharer was recorded as the same entity as the initial receiver.

\subsubsection{Inclusion of More Entities}
In \cite{Csomor23}, entities between the initial receiver and attacker who simply passed on data to one another were grouped together as a single entity.
At most one data handler was present, which might have represented multiple entities or transfers.
Rather than determining whether entities were relevant enough to include and losing relevant information due to grouping entities together (e.g., the length of a data flow, different parties involved), 
we adopted a structure where any parties identified by an article were included as entities in the data flow.
\clearpage
\onecolumn
\section{Data Collection Counts}
\label{sec:dc-counts}

\begin{table*}[h!]
\caption{Number of articles excluded and remaining after each step described in \Cref{sec:methods}, and the reasons for their exclusion. Some values are unavailable due to differences in the way we recorded data for different news sources. For WIRED and ABC, we only recorded information on articles that \emph{did not} meet any exclusion criteria for the format or title of the article (i.e. articles that passed the format and title check), so we cannot report data on articles that failed the format or title checks. For The Times, information on which criterion was used to exclude a given article was sometimes recorded in a freeform manner that made it difficult to obtain per-criterion counts; we have omitted values that could not be easily counted. The values in the Total column are omitted where an accurate total cannot be given because that value is missing for an individual news source.}
\centering
\begin{tabular}{llccccccc}

 & & \rotatebox[origin=l]{90}{The New York Times} & \rotatebox[origin=l]{90}{WIRED} & \rotatebox[origin=l]{90}{The Times} &  \rotatebox[origin=l]{90}{The Register} &  \rotatebox[origin=l]{90}{The Sydney Morning Herald} &  \rotatebox[origin=l]{90}{ABC} &  \rotatebox[origin=l]{90}{Total}\\
\toprule

\multicolumn{2}{l}{Articles Returned by Search} & 1858 & - & 978 & 649 & 941 & - & - \\
\midrule

\multirow{5}{0.6em}{\rotatebox[origin=c]{90}{Format}}
  & Not Text-Based & 42 & - & 13 & 4 & 0 & - & - \\
  & Not News Site Proper & 98 & - & 0 & 0 & 0 & - & - \\
  & Not Standard Format & 239 & - & 83 & 10 & 19 & - & - \\
  & Opinion Category & 145 & - & 47 & 16 & 98 & - & - \\
\cmidrule{3-9}
  & Remaining After Format Check & 1334 & - & 835 & 619 & 824 & - & - \\
\midrule

\multirow{5}{0.6em}{\rotatebox[origin=c]{90}{Title}}
  & Digital Security Incident & 11 & - & - & 77 & 24 & - & - \\
  & Product Recommendation & 84 & - & - & 73 & 8 & - & - \\
  & Tutorial & 8 & - & - & 1 & 0 & - & - \\
  & Real Estate and Travel & 32 & - & - & 0 & 87 & - & - \\
\cmidrule{3-9}
  & Remaining After Title Check & 1199 & 358 & - & 468 & 705 & 117 & -\\
\midrule

\multirow{8}{0.6em}{\rotatebox[origin=c]{90}{Contents}}
  & Digital Security Incident & 8 & 5 & - & 32 & 13 & 10 & - \\
  & Product Recommendation    & 11 & 0 & 4 & 6 & 19 & 1 & 41 \\
  & Tutorial                  & 1 & 0 & 0 & 0 & 0 & 3 & 4 \\
  & Real Estate and Travel    & 5 & 0 & 0 & 0 & 0 & 0 & 5 \\
  & Opinionated Article       & 0 & 5 & 0 & 0 & 9 & 4 & 18 \\
  & Duplicate                 & 13 & 5 & 6 & 4 & 7 & 4 & 39 \\
  & Inclusion Criteria Missing & 1077 & 302 & - & 405 & 633 & 80 & - \\
\cmidrule{3-9}
  & Remaining After Contents Check & 84 & 41 & 27 & 21 & 24 & 15 & 212 \\
\midrule

  & Excluded During Coding & 13 & 12 & 4 & 1 & 6 & 0 & 36 \\
  & Included During Coding & 0 & 0 & 1 & 1 & 0 & 0 & 2 \\
\midrule

\multicolumn{2}{l}{Final Included Article Count} & 71 & 29 & 24 & 21 & 18 & 15 & 178 \\

\bottomrule
\end{tabular}
\label{fig:numofarticlescoll-detailed}
\end{table*}

\twocolumn
\clearpage
\onecolumn
\section{Codebook}
\label{app:codebook}

\begin{longtable}{@{}l p{4.5cm} p{9cm}@{}}
\label{tab:codebook}\\
\toprule 
\textbf{Category} & \textbf{Code} & \textbf{Description}\\
\midrule
\endfirsthead
\toprule
\textbf{Category} & \textbf{Code} & \textbf{Description}\\
\midrule
\endhead

\multirow{2}{*}{Data Subject} & Person/People of Notoriety & Entity of broad recognition within one or multiple communities \\
& Person/People & Not a person of notoriety as defined above \\
\midrule

\multirow{7}{*}{Data Subject Behavior} & Noticeable Actions & Data subject does something noticeable, e.g., performing onstage, having a criminal record, going to a protest \\
& Actions are noticeable but routine for data subject & Data subject does something noticeable by ordinary standards, but is `everyday' for them \\
& Everyday online interactions & Data subject uses internet and devices as expected \\
& Everyday offline interactions & Data subject behaves inconspicuously \\
& Unknown & Not enough information to determine \\
\midrule

\multirow{12}{*}{Data Handler Type} & Same as Left & This data handler is the same as the previous entity in the data flow \\
& Individual(s) & An individual person \\
& Individual(s) - Personal Connection & Individual who is related to or knows the data subject \\
& Organization & An organized group of people \\
& Organization - Business & A commercial business \\
& Organization - Other & Another group \\
& Government & A governing body of a state, a government department, or an entity with equivalent access and resources\\
& Government - Law Enforcement & Law enforcement agents working for a government \\
& Government - Individual Agent & An agent of government or law enforcement acting in their capacity as an individual, not on behalf of the larger body \\
& Unknown - Type & Not enough information to determine what type of entity \\
& Unknown - Which Entity & Not enough information to determine which entity the IR or IS was \\
\midrule

\multirow{5}{*}{Data Access Type} & Access & Data is accessed: the active party is the receiver \\
& Access - Legal & Data is accessed through legal system, e.g., court orders, warrants, or a request from law enforcement \\
& Access - Physical & Data obtained through direct physical proximity to data subject \\
& Access - Technological & Data is not publicly available, but is obtained by using technology \\
\midrule

\multirow{2}{*}{Data Sharing Type} & Sharing & Data is shared: the active party is the sharer \\
& Sharing - Technological & Data is shared using technology \\
\midrule

\multirow{8}{*}{Data Transfer Type} & Transfer & Neither sender or receiver (or both sender and receiver) are active parties in the data transfer \\
& Transfer - Financial & Data is accessed through monetary means, e.g., paying a service to use an SDK \\
& Transfer - Technological & Data is transferred using technology \\
& Unknown & Not enough information to determine which parties are active in the transfer \\
& Preexisting & No need to create access, e.g., family DNA \\
& No Data Flow & Used when the left/right parties are the same \\
\midrule

\multirow{5}{*}{Initial Transfer Consent} & Access - Unknowing & Data subject is not aware data is accessed \\
& Access - Uncontrolled & Data subject is aware data is accessed but has no control over the access, i.e., no consent mechanism exists (often CCTV, mugshots) \\
& Sharing - Required for Service & Data subject is aware data is shared and the data is required by a service used by the data subject \\
& Sharing - Under Pressure/Coercion & Data subject is aware data is shared and allows sharing under some form of pressure or under false pretenses \\
& Sharing - Voluntary & Data subject is aware data is shared and allows sharing without pressure \\
& Unknown & Not enough information to determine \\
& No Data Flow & Used when the left/right parties are the same \\
\midrule
\ \\
\ \\
\multirow{7}{*}{Privacy Violation} & Collection of Data & Data is collected outside of accepted practices \\
& Disclosure of Data & Data is disclosed outside of accepted practices \\
& Disclosure of Data - To Third Party & Unaccepted disclosure is restricted to specific parties \\
& Disclosure of Data - To Public & Unaccepted disclosure is not restricted, e.g., posted publicly on the internet \\
& Unspecified Transfer of Data & Unclear whether it's a collection or disclosure, or something that is both (e.g., Transfer - Financial) \\
& Use of Data & Data is used outside of accepted practices \\
\midrule

\multirow{7}{*}{Targeting} & Targeted & The data subject is selected as the target of the attack \\
& Targeted - Indirect & The data subject is selected as the target of the attack by an earlier party through whom the data is passed to the attacker \\
& Subgroup & Data is filtered to select a subset of people with a shared attribute as aim of attack \\
& Untargeted & No or very little aim is employed for attack \\
& Unknown & Not enough information to determine \\
\midrule

\multirow{14}{*}{Data Type} & Identification & Unique identifiers of an individual, e.g., name, email address, phone number, home address \\
& Location & Location history, path data, or current location \\
& Usage & Data from usage of a device or app, e.g., tracking, behavioral data \\
& Technical & Data identifying a user's device and/or online profile, e.g., IP address, advertising IDs \\
& Visual & Pictures and/or videos \\
& Audio & Audio recordings \\
& Biometric & Measurements of physical characteristics used for verifying the identity of individuals, e.g., fingerprints, DNA, facial recognition data \\
& User Text & User-generated textual content, e.g., text messages, posts on social media \\
& Demographic & Attributes of human populations, e.g., age, income, gender, sexuality \\
& Other & Another kind of data \\
& Unknown & Not enough information to determine \\
\midrule

\multirow{10}{*}{Harm} & Psychological & Harms that result in emotional distress or disturbance \\
& Psychological - Threats & The attacker threatens the data subject \\
& Psychological - Harassment & The attacker harasses the data subject \\
& Psychological - Stalking & The attacker stalks the data subject \\
& Economic & Monetary loss or a loss in the value of something \\
& Economic - Job Loss & The data subject loses their job, e.g., quit, termination \\
& Economic - Job Prospects & Increased difficulty in finding a job or devaluation of skills \\
& Discrimination & Entrenching inequality and disadvantaging people based on characteristics or affiliations \\
& Physical & Harms that result in bodily injury or death \\
& Physical - Attack or Injury & Bodily injury \\
& Physical - Death & Death, e.g., murder, suicide \\
& Physical - Arrest & Arrest by law enforcement \\
& Reputational & Injuries to an individual's reputation and standing in a community, e.g., defamation \\
& Reputational - Legal prosecution & Legal investigation or conviction \\
& Autonomy & Restricting, undermining, inhibiting, or unduly influencing people's choices \\
& Autonomy - Targeted ads & Advertisement personalized to the data subject through the use of their personal information \\
& Non-Consensual Explicit Imagery (NCEI) & The creation or spread of non-consensual explicit imagery or similar materials \\
& Mass surveillance & Any systematic attention to a person's life aimed at exerting influence over it, performed over large (city-wide), otherwise anonymous publics \\
\midrule
\ \\
\multirow{4}{*}{Legality} & Legal penalty & Legal penalty against the initiator of privacy violation or attacker, e.g., fines, lawsuit, arrest \\
& Ongoing legal action & Unresolved legal action that may or may not lead to a penalty, e.g., an ongoing lawsuit or investigation \\
& Unknown & Not enough information in article to determine \\
\midrule

\multirow{25}{*}{Technology} & AI / Artificial Intelligence & Technology that performs functions typically thought of as requiring human intelligence, such as reasoning, recognizing patterns or understanding natural language. \footnote{This definition is modified from the AI Incident Database \cite{aiid}} \\
& AI - ML / Machine Learning & - \\
& AI - ML - LLM / Large Language Models & ChatGPT, Bing Copilot, Google Gemini, Character.AI \\
& AI - ML - FR / Facial Recognition & - \\
& AI - ML - Image Generation & Nudification apps, image generators (e.g., DALL-E) \\
& Communications & Technologies, platforms, or platform features intended for one-to-one communication \\
& News Media & - \\
& Image/Video Capture & Cameras \\
& Image/Video Capture - Mobile phone cameras & - \\
& Image/Video Capture - Surveillance drones & - \\
& Image/Video Capture - Surveillance cameras & - \\
& Spyware & Software installed on a device to monitor the activity of a user \\
& Smart devices & Smart TVs, smart refrigerators \\
& Online Platform & Any unspecified online application \\
& Online Platform - Social Media & Platforms or platform features designed for building a public profile, sharing user content, often `friending'/`following' other users, and meeting new people \\
& Online Platform - Websites & Platform specifically identified as a website \\
& Online Platform - Mobile apps & Platform specifically identified as a mobile app (a program built to run on a mobile operating system, e.g., Android or iOS) \\
& Bossware & Online platforms that employers use to monitor employees \\
& Tracking Device & A device that is built specifically to track a real-time location, e.g., GPS trackers, AirTags \\
& Other / Non-specific \\
\midrule

\multirow{5}{*}{Attacker Motivation} & Intentional & A goal is reached through purposefully causing harm, or the harm itself is the goal \\
& Intentional - Financial Gain & Harm is caused or knowingly accepted for profit \\
& Collateral & Harm is caused as an unintended result of a different goal \\
& Unknown & Not enough information to determine \\
\midrule

\multirow{5}{*}{Attacker Power} & High & Attack requires the resources (i.e., power, money) of a government actor or significant corporate entity \\
& Medium & Attack requires power not at the level of a government or significant corporate entity, but above that of entity being attacked \\
& Low & Attack requires power of an average person \\
& Unknown & Not enough information to determine \\
\bottomrule
\end{longtable}

\twocolumn
\section{Details on Application of Codes}
\label{sec:code-details}
Some patterns of events recurred frequently enough that we established guidelines for handling them during the coding process.

\paragraph{Data Subject and Initial Sharer}
As described in \Cref{sec:role-action}, in every data flow either the data subject and initial sharer are the same entity or the initial sharer and initial receiver are the same entity (depending on the data subject's involvement in the initial transfer of data).
We frequently encountered cases where a data subject used an online service that collected their data for tracking/advertising purposes.
Generally we determined the data subject was meaningfully involved in the first transfer of data (and so the data subject and initial sharer were the same entity), because individuals are broadly are aware of some degree of online tracking.
We made exceptions to this guideline when there was indication in the article that data subjects were unaware of or surprised by the tracking, or when the data subjects were children.



\paragraph{Social Media}
Articles often refer to ``social media'' without specifying a particular platform (e.g., to say that information had proliferated over social media).
In these cases, despite not knowing exactly which platforms were involved, we assumed they included one or more large, well-known social media platforms 
(e.g., Facebook, Instagram, TikTok).
Therefore, when relevant, we coded the type of such entities as `Organization - Business' and their power as `High'.

\paragraph{Data Brokers}
Because data brokers are primarily involved in buying and selling data, when an article specified that data was transferred to/from a data broker, we assumed those transfers were of a financial nature and coded them as `Transfer - Financial', unless there was some indication otherwise.


\paragraph{Targeted Advertising}
Targeted ads are coded as `Untargeted' because, while the content of the advertising is targeted by user attributes, ads are typically served to all users of a given platform.
Because the purpose of targeted advertising is generally financial (e.g., to increase sales or to receive money from advertisers), when the harm of an attack was targeted advertising, we coded the motivation of the attack as `Intentional - Financial Gain' (unless we had some indication otherwise).

\paragraph{Initiator of Privacy Violation}
Typically, when the privacy violation is a \emph{sharing}, the initiator of the privacy violation is the entity sharing, and when the privacy violation is an \emph{access}, then the initiator 
is the receiving entity.
If the privacy violation is a \emph{data transfer}, then the initiator is often assigned to both parties involved, because they are both or neither responsible for initiating the transfer.
We made exceptions for cases where the sharing was performed without consent (e.g., a transfer coded as `Sharing Under Pressure/Coercion').
In such cases, we identified the entity receiving the data as the initiator of the privacy violation.


\paragraph{Identifying the Privacy Violation}
Often, multiple events in a data flow were identifiable from the article as `outside the scope of accepted practices'.
Based on the definition of a privacy violation we used for our analysis (Def.~\ref{def:privviol}) we considered only the first such event to be the privacy violation.
Our rationale was that an entity only \emph{loses control} of their information once, although further later events involving the information may also be outside the scope of accepted practices.

\paragraph{Biometric Data}
Any image showing a person's face might be considered biometric data.
For the purposes of our coding we deferred to the narrative of the article, and only used the `Biometric' code when the data involved was specifically described as ``biometric'' by the article, or if the data was being used for facial recognition purpose.

\paragraph{Image/Video Capture}
By inference, whenever image/video data is involved in a data flow, some form of image/video capture technology must have been involved.
However, we chose to only include the `Image/Video Capture' code when the process of taking an image/video was relevant to the case or described in the article.

\paragraph{Online Platforms}
There is a wide variety of platforms and services that fall under the umbrella of the `Online Platform' code, and many of them have both websites and mobile apps.
Often an article would describe the usage of one of these platforms without specifying what type of software was involved.
We used the `Online Platform' tag when some ``app'', ``service'', or ``platform'' was involved.
We specified `Online Platform - Mobile Apps' or `Online Platform - Websites' when there was clear indication that the software involved in the data flow was a mobile app or website.

\paragraph{Communications and Social Media}
The `Communications' and `Online Platform - Social Media' codes capture similar technologies.
We used `Communications' when the technology described was focused on communications with people one already knows (e.g., SMS, email, Telegram, or Signal).
We used `Online Platform - Social Media' when the focus was on meeting new people or broadcasting information (e.g., Instagram, Facebook, dating apps).
In some cases we coded the same platform with different codes depending on how the platform was being used in the data flow (e.g., Instagram can facilitate both one-to-one direct messaging and posting information for a broad audience).
\clearpage
\onecolumn
\section{Visualization of All Data Flows}
\label{sec:data-vis}

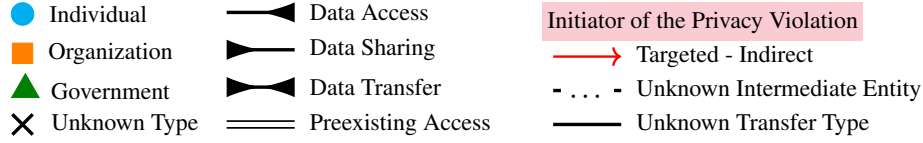
\begin{figure}[ht]
    \centering
    \caption{Legend for Data Flow Visualization}
    \label{fig:data-flow-legend}
    \scalebox{0.9}{%
    \begin{tikzpicture}[>=Latex]
        \def\hdist{0.5}
        \def\vdist{0.5}
        \tikzset{label distance=1.5mm}

        \node[rectangle, align=left, below=\vdist*3 of IR, xshift=14mm] (legend) {
            \begin{tikzpicture}[>=Latex]
                \tikzset{label distance=1mm}

                \node [person, label=right:{Individual}] (indiv) {};

                \node [orga, below=\vdist*0.5 of indiv, label=right:{Organization}] (orga) {};

                \node [gov, below=\vdist*0.8 of orga, label=below:{Government},yshift=4pt] (gov) {};

                \coordinate (unk) at (0, -1.8);
                \draw[line width=0.5mm, color=black] (unk) ++(-0.15, -0.15) -- ++(0.3, 0.3);
                \draw[line width=0.5mm, color=black] (unk) ++(-0.15, 0.15) -- ++(0.3, -0.3);
                \node at (0.3, -1.8) [right] {Unknown Type};

                \draw[-{<[length=5mm,width=3mm]}, line width=0.5mm] (3, -0.15) -- (4, -0.15);
                \node [right=2mm] at (3.9, -0.15) {Data Access};

                \draw[{>[length=5mm,width=3mm]}-, line width=0.5mm] (3, -0.7) -- (4, -0.7);
                \node [right=2mm] at (3.9, -0.7) {Data Sharing};

                \draw[{>[length=5mm,width=3mm]}-{<[length=5mm,width=3mm]}, line width=0.5mm] (3, -1.25) -- (4, -1.25);
                \node [right=2mm] at (3.9, -1.25) {Data Transfer};

                \draw[-, line width=0.25mm] (3, -1.75) -- (4, -1.75);
                \draw[-, line width=0.25mm] (3, -1.85) -- (4, -1.85);
                \node [right=2mm] at (3.9, -1.8) {Preexisting Access};

                \draw[-, line width=0.5mm] (7.8, -1.8) -- (8.8, -1.8);
                \node [right=2mm] at (8.7, -1.8) {Unknown Transfer Type};

                \draw[-,line width=0.5mm] (7.8,-1.3) -- (8.8,-1.3) node[midway, fill=white, yshift=4pt] {\scalebox{1.3}{$\cdots$}};
                \node [right=2mm] at (8.7, -1.3) {Unknown Intermediate Entity};

                \node[fill=tudred!20, rectangle, inner sep=2pt,align=center] at (10, 0) {\vphantom{y}\strut Initiator of the Privacy Violation};

                \draw[{-Computer Modern Rightarrow},line width=0.35mm,draw=red] (7.8,-0.8) -- (8.8,-0.8);
                \node [right=2mm] at (8.7, -0.8) {Targeted - Indirect};

                \node[below=\vdist of unk, xshift=20em] {Labels of the form XXX123 refer to corresponding labels in the dataset};
            \end{tikzpicture}
        };
    \end{tikzpicture}
    }
\end{figure}

\centering
\includegraphics[scale=1.05]{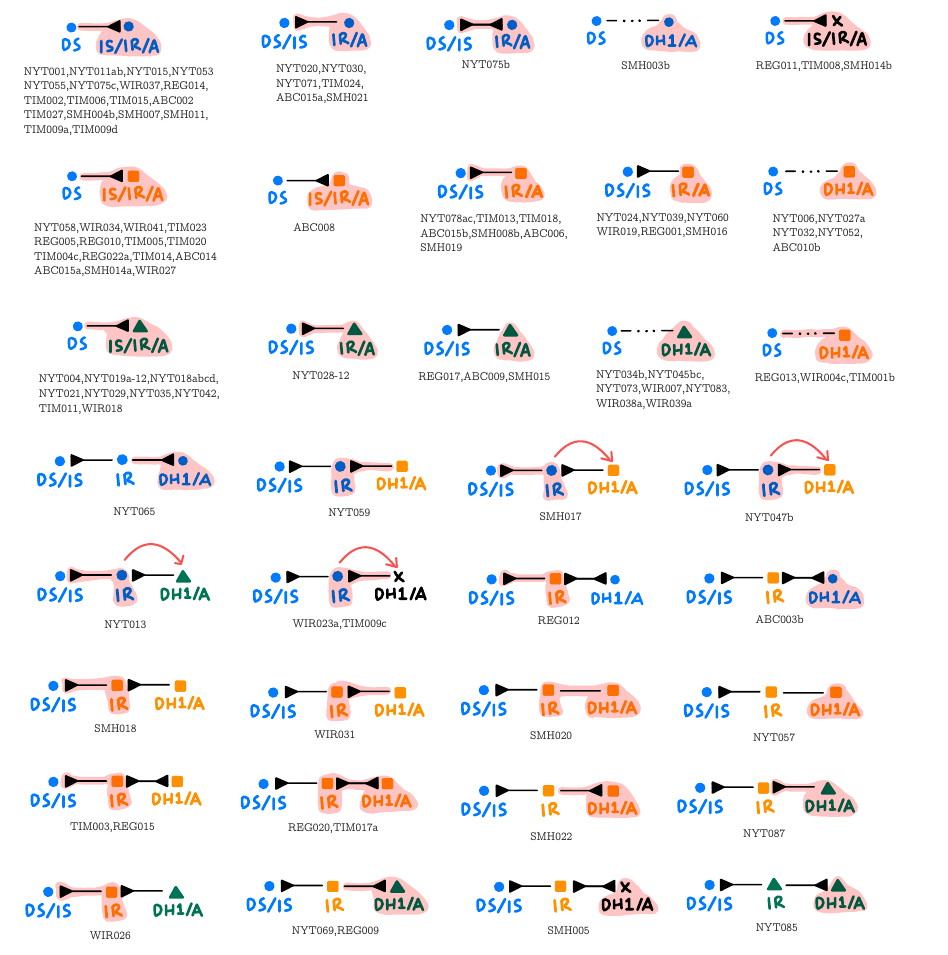}

\includegraphics[scale=1.1]{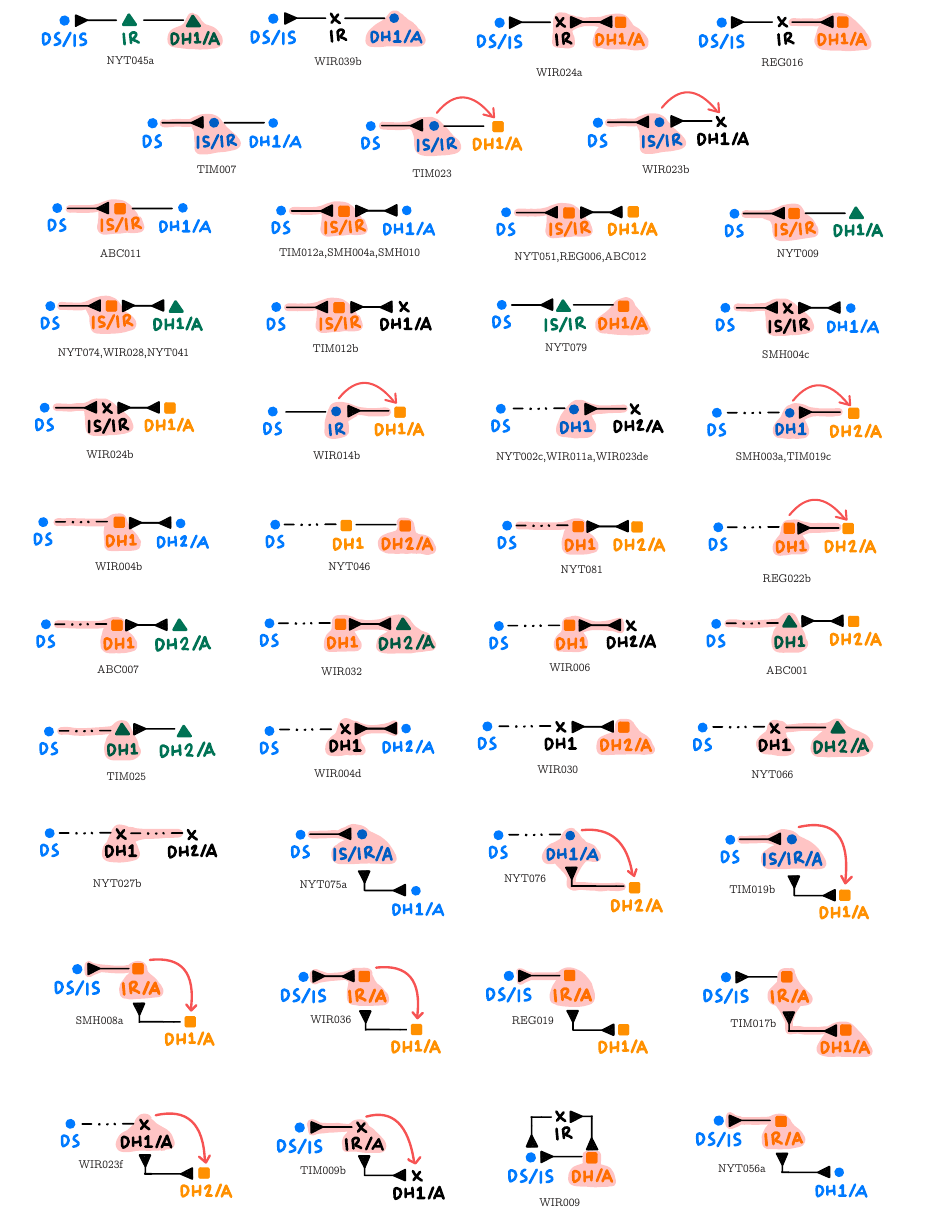}

\includegraphics[scale=1.1]{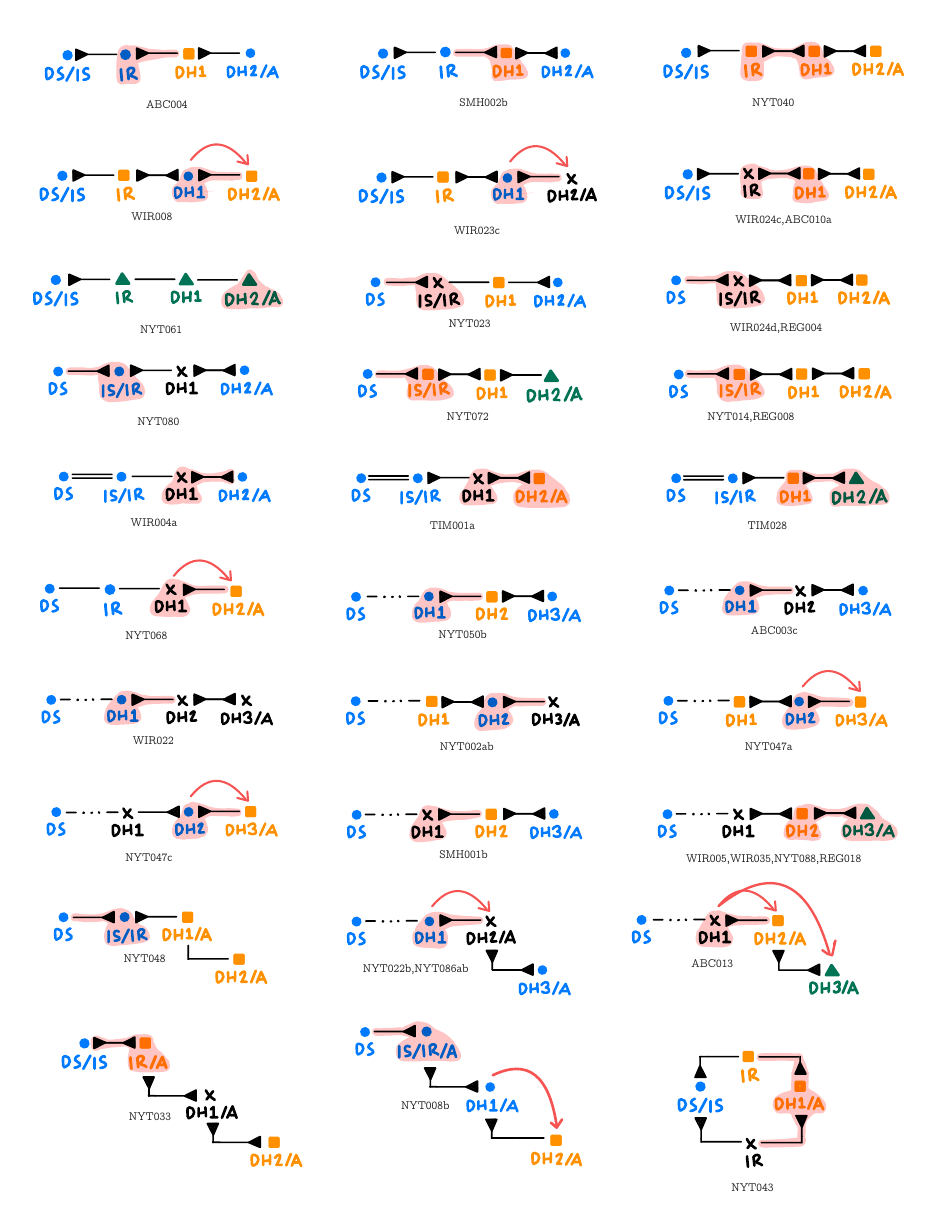}

\includegraphics[scale=1.1]{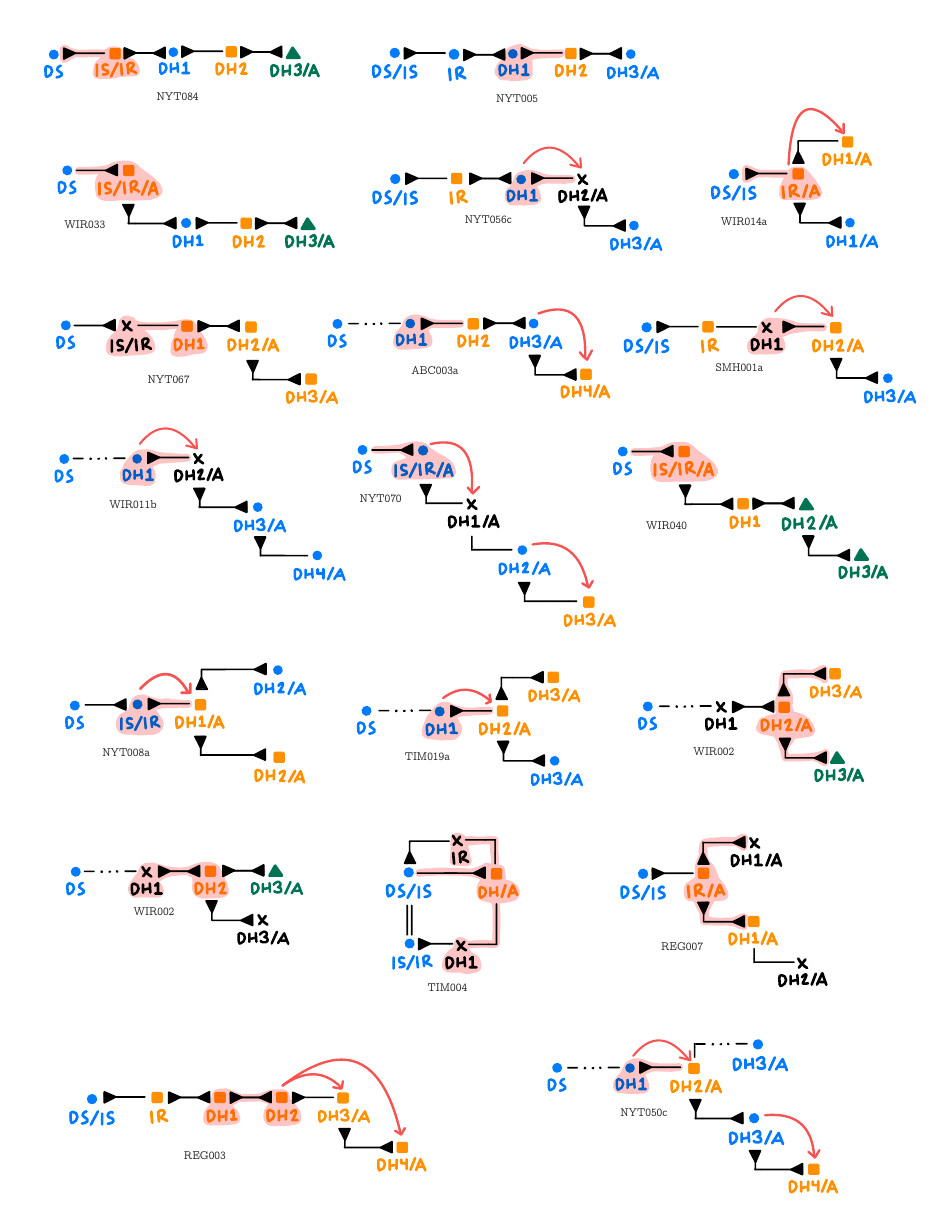}

\includegraphics[scale=1.1]{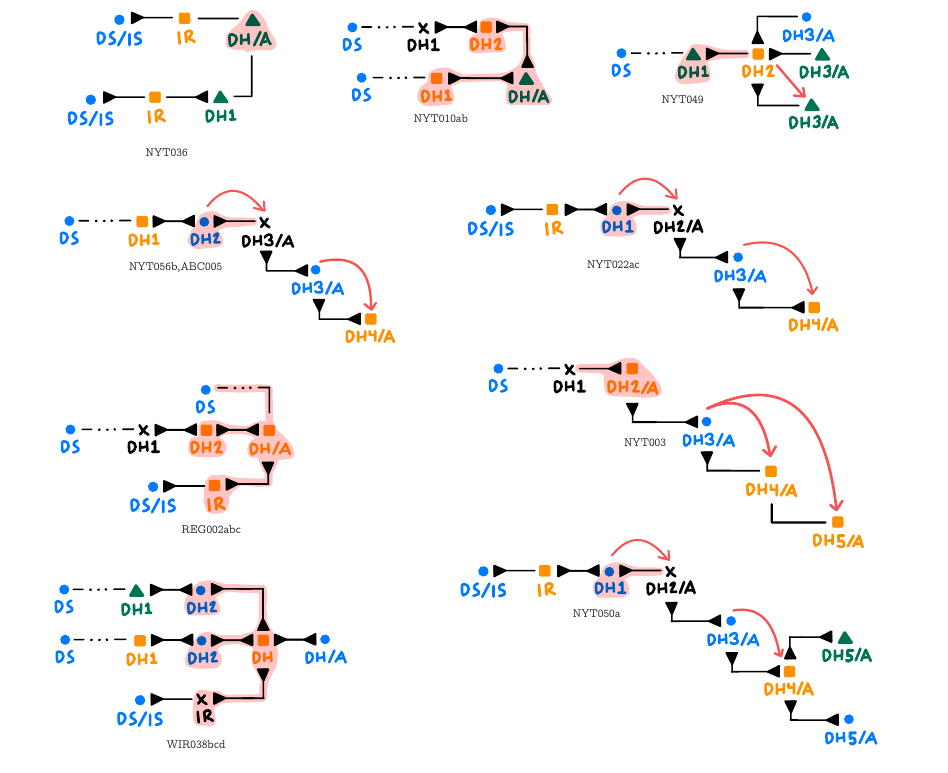}
\clearpage
\section{Results of Article Analysis}
\label{sec:heatmap}
\ifacmversion\else\vspace{-20pt}\fi

\par\smallskip
\noindent
\begin{table}[h]
\centering
   \caption{High level categories of harms that we identified along with harms that fall under each harm category.} \label{tbl:highlevelharmslist}
\begin{tabular}{llll}
\toprule
Psychological & Physical & Economic & Reputational \\
\rotatebox[origin=c]{180}{$\Lsh$} Threats & \rotatebox[origin=c]{180}{$\Lsh$} Attack or Injury & \rotatebox[origin=c]{180}{$\Lsh$} Job Loss & \rotatebox[origin=c]{180}{$\Lsh$} Legal Prosecution \\
\rotatebox[origin=c]{180}{$\Lsh$} Harassment & \rotatebox[origin=c]{180}{$\Lsh$} Death & \rotatebox[origin=c]{180}{$\Lsh$} Job Prospects\\
\rotatebox[origin=c]{180}{$\Lsh$} Stalking & \rotatebox[origin=c]{180}{$\Lsh$} Arrest& \\
 & &  \\
Autonomy & Discrimination &  NCEI & Mass Surveillance \\
\rotatebox[origin=c]{180}{$\Lsh$} Targeted Ads \\
\bottomrule
\end{tabular}
\end{table}
\par\smallskip

\begin{table}[h]
\caption{Number of attacks that caused each harm. 
Attacks that cause multiple harms are counted once for each harm.
Subcodes listed under a parent code are denoted by an arrow.
}
\centering
\scalebox{0.9}{%
\begin{tabular}{lc|lc}\toprule
Harm & Attacks & Harm & Attacks \\
\midrule
Autonomy & 19 & NCEI & 108 \\
\rotatebox[origin=c]{180}{$\Lsh$} Targeted Ads & 37 & Physical & 18 \\
Discrimination & 9 & \rotatebox[origin=c]{180}{$\Lsh$} Arrest & 10 \\
Economic & 33 & Psychological & 135 \\
Mass Surveillance & 26 & Reputational & 31 \\
\bottomrule
\end{tabular}
}
\label{fig:numharms}
\end{table}

\begin{figure}[h]
\includegraphics[scale=1]{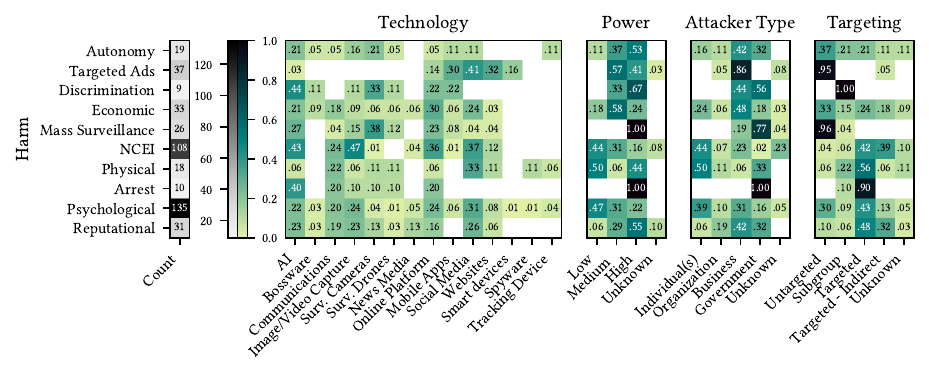}
\caption[Relative Occurrences]{Frequency of co-occurrence of a particular code for Technology, Targeting, or Power with a particular code for Harm. Multiple attacks with multiple different harms, associated technologies, and other attributes may stem from the same privacy incident. One cell in this figure displays the number of attacks which both caused a particular harm (left axis) and involved another particular code (bottom axis), divided by the total count of attacks which caused that harm (leftmost column). 
}
\label{fig:relative-occurrences}
\end{figure}

\clearpage
\section{System Intervention Results}
\label{sec:interventions}

\ifacmversion\else\vspace{-15pt}\fi
\begin{longtable}{@{}p{5.5cm} p{3.5cm} p{3.2cm} p{4.2cm}@{}}
\caption{Examples of System Interventions}
\label{tab:interventions} \\
\toprule 
\textbf{System Description} & \textbf{Intervention} & \textbf{Type} & \textbf{Restrictions}\\
\midrule
\endfirsthead
\toprule
\textbf{System Description} & \textbf{Intervention} & \textbf{Type} & \textbf{Restrictions}\\
\midrule
\endhead

Automatic Content Recognition used to target ads on Smart TVs [Tab.~\ref{tab:article-refs}, Row~\ref{r17}] & Opt-out mechanisms \cite{IMC:AVDCMS24} & User-based regulatory & None (intervention does not apply in secondary analysis where financial gain is considered) \\

Data collection on public transport network causes psych.~harm [Tab.~\ref{tab:article-refs}, Row~\ref{r9}] & Turn Wi-Fi off on device & User-based regulatory & Partial restriction on user: freedoms during commute; Partial restriction on service: collects less information \\

Addressed leaked from Snap Map used for blackmail [Tab.~\ref{tab:article-refs}, Row~\ref{r8}] & Spoof device location & User-based tech & Partial restriction on user: use of Snapchat features \\

Facial recognition (FR) in vending machine causes psych.~harm [Tab.~\ref{tab:article-refs}, Row~\ref{r20}] & Accessories that impede FR \cite{CCS:SBBR16,hyperface} & User-based tech & None \\

Tracking Covid spread and enforcing quarantine requirements [Tab.~\ref{tab:article-refs}, Row~\ref{r7}] & Privacy-preserving contact tracing \cite{dp3t} & System-based privacy engineering & Partial restriction on service: less control of individuals and ability to enforce rules \\

Partner accessing messages and using as evidence [Tab.~\ref{tab:article-refs}, Row~\ref{r21}] & Refined/hidden authentication mechanisms \cite{CHI:FPMLRD16} & System-based privacy engineering & None \\

Driving behavior applications providing data to third parties, leading to increase in insurance premiums [Tab.~\ref{tab:article-refs}, Row~\ref{r1}] & Not share data with third parties & System-based regulatory & None (intervention does not apply when financial gain is considered) \\

Facial recognition use in TSA pre-security checks causes psych.~harm [Tab.~\ref{tab:article-refs}, Row~\ref{r3}] & Not use facial recognition tech (revert to paper-based methods) & System-based regulatory & Partial restriction on service: less efficient processing \\
\bottomrule
\end{longtable}

\ifacmversion\else\vspace{-15pt}\fi

\begin{figure*}[ht!]
\includegraphics[scale=0.35]{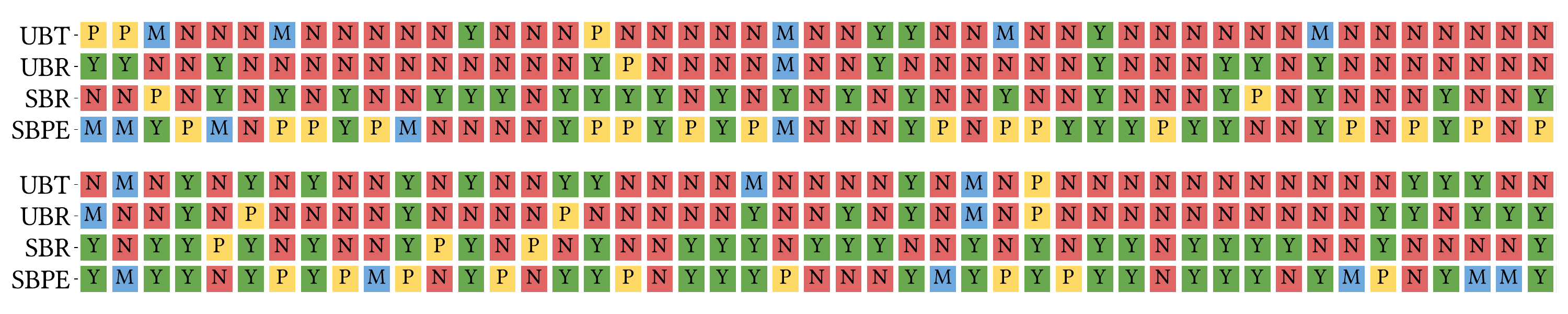}
\caption[System Intervention Results]{%
System Intervention Results.  Each column of 4 cells corresponds to one analyzed system and whether each intervention type (UBT = User-based tech, UBR = User-based regulatory, SBR = System-based regulatory, SBPE = System-based privacy engineering) applies. \tikz[baseline=(square.base)]{
  \node[draw=green!60!olive!40, fill=green!60!olive!40, rectangle, inner sep=1pt] (square) {Y};
} = Yes,
\tikz[baseline=(square.base)]{
  \node[draw=red!50, fill=red!50, rectangle, inner sep=1pt] (square) {N};
}
= No,
\tikz[baseline=(square.base)]{
  \node[draw=orange!30!yellow!50, fill=orange!30!yellow!50, rectangle, inner sep=1pt] (square) {P};
}
= Partial,
\tikz[baseline=(square.base)]{
  \node[draw=cyan!45, fill=cyan!45, rectangle, inner sep=1pt] (square) {M};
}
= Maybe}
\label{fig:interventions-all}
\end{figure*}


\begin{minipage}{0.5\textwidth}
    \centering
    \includegraphics[scale=0.4]{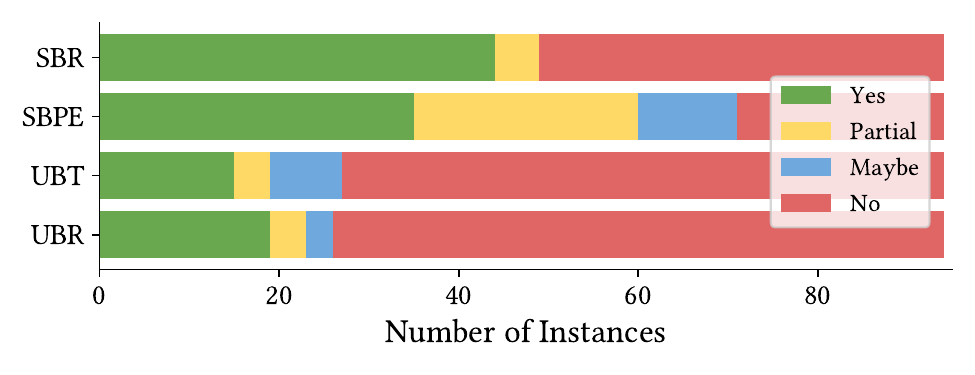}
    \vspace{-10pt}
    \captionof{figure}{Number of cases where each intervention type applies. SBX = System-based X, UBX = User-based X; R = (Self-) Regulatory, T = Technical, PE = Privacy Engineering}
    \label{fig:intervention-counts}
\end{minipage}%
\begin{minipage}{0.5\textwidth}
    \centering
    \includegraphics[scale=0.31]{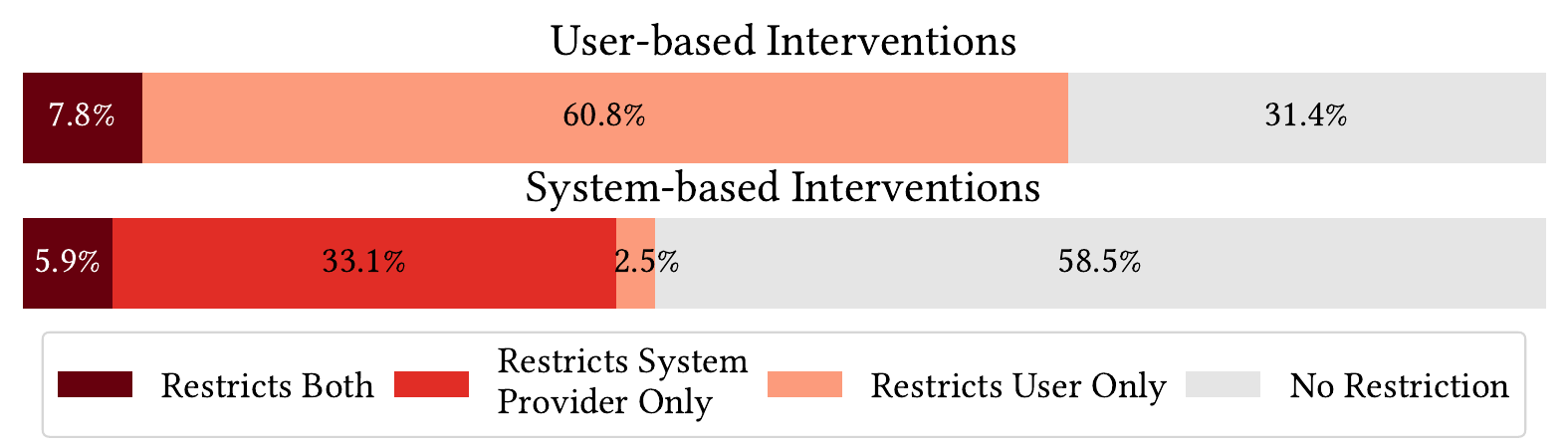}
    \captionof{figure}{Proportion of Interventions with Restrictions}
    \label{fig:restriction-counts}
\end{minipage}

\clearpage

\newcounter{rownumber}[figure] 
\setcounter{rownumber}{0}

\section{Referenced Articles}
\label{sec:article-refs}
\begin{longtable}{@{}l l l p{10cm} p{3cm}@{}}
\caption{Referenced Articles}
\label{tab:article-refs}\\
\toprule 
& \textbf{Ref.} & \textbf{Source} & \textbf{Title} & \textbf{Date Published}\\
\midrule
\endfirsthead
\toprule
& \textbf{Ref.} & \textbf{Source} & \textbf{Title} & \textbf{Date Published}\\
\midrule
\endhead

\refstepcounter{rownumber}\label{r1}1 & Ex.~\ref{ex:nyt014-40} & NYT & Florida Man Sues G.M. and LexisNexis Over Sale of His Cadillac Data & March 14, 2024 \\
\refstepcounter{rownumber}\label{r2}2 & Ex.~\ref{ex:nyt014-40} & NYT & Is Your Driving Being Secretly Scored? & June 9, 2024 \\
\refstepcounter{rownumber}\label{r3}3 & \Cref{sec:interventions} & NYT & Senators Seek to Curb Facial Recognition at Airports, Citing Privacy Concerns & May 7, 2024 \\
\refstepcounter{rownumber}\label{r4}4 & Ex.~\ref{ex:tiktok} & NYT & U.S. Sues TikTok Over Child Privacy Violations & August 2, 2024 \\ 
\refstepcounter{rownumber}\label{r5}5 & Ex.~\ref{ex:schools} & NYT & Spying on Student Devices, Schools Aim to Intercept Self-Harm Before It Happens & December 9, 2024 \\
\refstepcounter{rownumber}\label{r6}6 & \Cref{sec:interv-methods} & NYT & H.I.V. Groups Warn of Privacy Risks in How C.D.C. Tracks Virus Samples & February 9, 2024 \\
\refstepcounter{rownumber}\label{r7}7 & \Cref{sec:interventions} & NYT & Xi Jinping's Recipe for Total Control: An Army of Eyes and Ears & May 25, 2024 \\
\refstepcounter{rownumber}\label{r8}8 & Ex.~\ref{ex:ubereats} & ABC & Young people share their experiences of being doxxed — from 50 food deliveries a day to being `terrified' & May 15, 2024 \\
\refstepcounter{rownumber}\label{r9}9 & \Cref{sec:interventions} & ABC & Privacy fears over Sydney light rail services pinging data from passengers' phones & November 17, 2024 \\
\refstepcounter{rownumber}\label{r10}10 & & REG & Study: Thousands of businesses just love handing over your info to Facebook & January 18, 2024 \\
\refstepcounter{rownumber}\label{r11}11 &  & REG & `Scandal-plagued' data broker tracked visits to `600 Planned Parenthood locations' & February 15, 2024 \\
\refstepcounter{rownumber}\label{r12}12 & \Cref{sec:purpose-method} & REG & eBay to cough up \$3M after cyber-stalking couple who dared criticize the souk & January 11, 2024 \\
\refstepcounter{rownumber}\label{r13}13 & \Cref{sec:purpose-method} & SMH & `I'm watching you bitch': How even the smart fridge is being used as a weapon of family violence & April 28, 2024 \\
\refstepcounter{rownumber}\label{r14}14 & \Cref{sec:purpose-method} & SMH & AirTags and GPS trackers: How domestic abusers stalk their victims & June 25, 2024 \\
\refstepcounter{rownumber}\label{r15}15 & \Cref{sec:purpose-method} & SMH & Millions use parental tracking apps but this psychologist isn't sold & July 8, 2024 \\
\refstepcounter{rownumber}\label{r16}16 & \Cref{sec:consent} & TIM & Think your phone is listening to you? That's proximity advertising & December 21, 2024 \\
\refstepcounter{rownumber}\label{r17}17 & \Cref{sec:consent} & TIM & You think you're watching smart TV \dots In fact it's watching you & November 11, 2024 \\
\refstepcounter{rownumber}\label{r18}18 & \Cref{sec:purpose-method} & TIM & `Stalkerware' generation set to think tracking partners is normal & February 5, 2024 \\
\refstepcounter{rownumber}\label{r19}19 & \Cref{sec:consent} & TIM & Britain's FBI could catch killers by tracking down relatives & February 24, 2024\\
\refstepcounter{rownumber}\label{r20}20 & \Cref{sec:interventions} & WIR & A Vending Machine Error Revealed Secret Face Recognition Tech & February 24, 2024 \\
\refstepcounter{rownumber}\label{r21}21 & \Cref{sec:interventions} & WIR & She Escaped an Abusive Marriage---Now She Helps Women Battle Cyber Harassment & December 5, 2024 \\
\bottomrule \\
\end{longtable}

\twocolumn

\end{document}